\documentclass[trackchanges]{aastex701}

\usepackage{xcolor}
\usepackage{amsmath}
\usepackage{svg}

\newcommand{\pzred}[1]{{#1}}

\begin{document}

\title{Radio Spectral Imaging and MHD Modeling of a CME-Driven Shock:\\ Connecting Solar Type II Radio Bursts with Shock-Surface Magnetic Geometry}

\author[0000-0001-6855-5799]{Peijin Zhang}
\affiliation{Center for Solar-Terrestrial Research, New Jersey Institute of Technology, Newark, NJ 07102, USA}
\email[show]{peijin.zhang@njit.edu}  

\author[orcid=0000-0002-2873-5688]{Weihao Liu}
\affiliation{Department of Climate and Space Sciences and Engineering, University of Michigan, Ann Arbor, MI 48109, USA}
\email[]{whliu@umich.edu}

\author[0000-0002-0660-3350]{Bin Chen}
\affiliation{Center for Solar-Terrestrial Research, New Jersey Institute of Technology, Newark, NJ 07102, USA}
\email[]{bin.chen@njit.edu}

\author[orcid=0000-0003-0472-9408]{Ward B. Manchester IV} 
\affiliation{Department of Climate and Space Sciences and Engineering, University of Michigan, Ann Arbor, MI 48109, USA}
\email{chipm@umich.edu}

\author[orcid=0000-0002-2325-5298]{Surajit Mondal}
\affiliation{National Centre for Radio Astrophysics, Tata Institute of Fundamental Research, Pune 411007, India}
\affiliation{Center for Solar-Terrestrial Research, New Jersey Institute of Technology, Newark, NJ 07102, USA}
\email{surajit.mondal@njit.edu}

\author[orcid=0000-0003-2872-2614]{Sijie Yu}
\affiliation{Center for Solar-Terrestrial Research, New Jersey Institute of Technology, Newark, NJ 07102, USA}
\email{sijie.yu@njit.edu}

\begin{abstract}

Solar type II radio bursts are widely regarded as signatures of shock waves propagating in the solar corona and are of particular importance for understanding shock-driven particle acceleration processes. 
Type II radio bursts often exhibit complex multi-lane and split-band features. The detailed spectral, temporal, and spatial structures carry key information about the shock properties and evolution. However, the physical origin of the multi-lane and split-band features remains unclear, largely due to a lack of spatially resolved data and understanding of the concurrent shock morphology and its magnetic-field context. 
In this work, we combine radio imaging spectroscopy of a multi-lane, split-band type II burst event with a three-dimensional global magnetohydrodynamic simulation of the associated coronal mass ejection-driven shock using the Alfv\'{e}n Wave Solar atmosphere Model-Realtime. \pzred{In this event, the burst intensity evolves from fundamental-emission dominated to harmonic-emission dominated. 
Meanwhile, the preferential emission source region moves from the Earth-facing side to the limb or far side, coinciding with quasi-perpendicular shock regions with enhanced Mach numbers. 
The observed spatial offset between the fundamental and harmonic sources is generally aligned with the projected shock-surface magnetic field from the simulation, consistent with anisotropic scattering in a magnetized turbulent plasma.}
These results establish a physical connection between type II radio sources and coronal shock magnetic geometry, providing new insight into the origin of the multi-lane features and their diagnostics of coronal shocks.

\end{abstract}


\section{Introduction} \label{sec:introduction}

Solar type II radio bursts are characterized by slowly drifting emission features in dynamic spectra and are commonly interpreted as plasma emission generated at coronal shock fronts \citep[e.g.,][]{NelsonMelrose1985}. \pzred{They commonly appear as fundamental--harmonic (F--H) pairs, and the relative intensity of the harmonic and fundamental bands ($I_H/I_F$) can differ from burst to burst and with heliographic longitude, reflecting coronal refraction, emission directivity, and viewing geometry \citep{jha2025relativeFH}.} These shocks are often associated with coronal mass ejections (CMEs) and play a central role in accelerating electrons and ions in the corona and heliosphere. Type II bursts frequently exhibit band-splitting, where two nearly parallel emission lanes appear simultaneously at similar frequencies. The band-splitting feature is often interpreted as emission from upstream and downstream regions of the shock \citep{Smerd1974, Vrsnak2002}, while an alternative scenario that involves spatially distinct source regions on different parts of the same shock front is also proposed  \citep{zhang2024AAshock}.

Despite their long-recognized diagnostic potential, the physical origin of the multi-lane and band-splitting features and its spatiotemporal relationship to shock geometry remain unclear. \pzred{The suprathermal electrons required for generating type II radio bursts are most commonly attributed to shock drift acceleration (SDA), in which particles gain energy through repeated reflection across the shock ramp \citep{holman1983sda, armstrong1985shock}. Although quasi-perpendicular geometry ($\theta_{Bn} \gtrsim 45^\circ$) is often regarded as more efficient for electron acceleration via the SDA mechanism, quasi-parallel shocks can also produce electrons energetic enough to generate type II emission when short large-amplitude magnetic structures (SLAMS) locally swing the field toward quasi-perpendicular geometry \citep{mann1994quasiparallel}.} Most previous studies rely primarily on total-power dynamic spectra, which provide no or limited constraints on the spatial structure and evolution of the radio sources. In particular, the role of the magnetic field orientation relative to the shock normal, usually described by the shock obliquity angle $\theta_{Bn}$, is known to be critical for particle acceleration efficiency but has remained difficult to evaluate observationally in the middle corona in the absence of spatially resolved radio data.

Recent advances in low-frequency radio interferometry, such as the Low Frequency Array (LOFAR), have enabled high-cadence imaging spectroscopy of the solar corona, allowing direct localization of radio burst emission \citep[e.g.,][]{Kontar2017, Normo2025AAshock,Morosan2019,zhang2024AAshock,morosan2025AAshocktypeII,zucca2025AAmultilane}. These observations have revealed that type II bursts can originate from spatially complex and evolving regions, which are not always consistent with simple radially expanding shock models.

The Owens Valley Radio Observatory's Long Wavelength Array (OVRO–LWA) is a low-frequency radio interferometer located in California, operating in the frequency range of 13–87 MHz with spectral imaging capability and a typical imaging dynamic range of 200:1 for the quiet Sun. The array consists of 352 dual-polarization dipole antennas arranged in a pseudo-random configuration, providing imaging capabilities with arcminute angular resolution at these frequencies (8 arcmin at 60 MHz). OVRO–LWA's wide field of view and high-cadence imaging spectroscopy make it particularly well suited for investigating dynamic solar radio phenomena, including type II and type III radio bursts. Previous OVRO–LWA studies have successfully localized type II radio sources in the corona and traced their spatial evolution, revealing complex source structures \citep[e.g.,][]{chen2025ApJlocmag}. These imaging studies have demonstrated the array's capability to track radio emission from CME-driven shocks and have provided new insight into the relationship between radio burst morphology and shock geometry.

In parallel, three-dimensional (3D) global magnetohydrodynamic (MHD) simulations have advanced significantly over the past decade, providing increasingly realistic descriptions of the solar corona, ambient solar wind, CME initiation, propagation, and shock formation \cite[e.g.,][and references therein]{chen2011coronal, reiss2023progress, corti2026advancing}. 
In particular, the Alfv\'{e}n Wave Solar atmosphere Model \citep[AWSoM;][]{sokolov2013magnetohydrodynamic, van2014alfven} and its extended version, AWSoM-Realtime \citep[AWSoM-R;][]{sokolov2021threaded}, self-consistently incorporate coronal heating, solar wind acceleration, and magnetic field evolution in 3D space, enabling quantitative comparisons with both \textit{in-situ} and remote-sensing observations of the ambient solar wind and CME evolution from the solar corona to the inner heliosphere \citep[e.g.,][]{sachdeva2019validation, shi2022awsom, zhao2024solar, manchester2025high, liu2025physics, liu2026testbed, liu2026simulating}. 

Within AWSoM(-R), a dedicated shock--capturing tool has recently been developed to identify the CME-driven shock front in MHD simulations and to systematically characterize the shock properties, such as the Mach number, density compression ratio, and shock obliquity angle \citep[see][]{liu2025physics, liu2026counterintuitive, chen2025evidence}. Together, these capabilities make global MHD simulation a powerful approach for investigating how shock geometry regulates the efficiency and spatial distribution of particle acceleration and for interpreting radio imaging observations. 

In this paper, we present OVRO–LWA imaging spectroscopy of a type II radio burst event with clear multi-lane and band splitting features, and compare the observed radio source evolution with the shock wave front captured in a 3D global MHD simulation by AWSoM-R. We demonstrate that the radio sources align with regions of quasi-perpendicular shock geometry, establishing a direct link between radio emission and coronal shock magnetic structure. 
In the following, Section \ref{section:obs2} describes the OVRO–LWA observations and data analysis methods; Section \ref{section:simu3} introduces the 3D MHD simulation and shock--capturing technique; Section \ref{section:surface4} presents the orientation of the shock surface relative to the coronal magnetic field and its relationship to the observed type II radio sources; and Section \ref{section:sum5} discusses and summarizes the implications for type II band splitting and shock-driven particle acceleration throughout this work.

\section{Observation} \label{section:obs2}

\pzred{The type II burst analyzed here occurred on 2024 November 18 during activity from NOAA active region AR~3901 near the east limb (S07E68). The region produced multiple M-class flares throughout that day, including M3.7 at 12:53~UT, M1.8 at 17:49~UT, and M2.0 at 19:15~UT. Southeast limb CMEs were reported in LASCO near 08:36 and 19:24~UT and were not Earth-directed. The SOHO/LASCO CME catalog \citep{gopalswamy2009soho} lists two entries with launch times near 17:45--17:58~UT and speeds of $\sim 380$--430~km~s$^{-1}$, likely associated with the 17:49~UT eruption. The OVRO--LWA type II emission below begins after $\sim$19:00~UT (Section~\ref{subsec:dynamic_spec}).}

\subsection{OVRO--LWA Observations and Data Processing}

The OVRO--LWA observations consist of both imaging spectroscopy and beamformed dynamic spectra. The beamformed dynamic spectra were obtained using 256 core antennas with a time resolution of 64 ms and a frequency resolution of 24~kHz, providing high temporal and spectral resolution for detailed analysis of the burst structure. Simultaneously, imaging observations were performed with an integration time of 10~s at a cadence of 20~s and a frequency resolution of 0.384~MHz, enabling spatial localization of the evolving radio sources.
Both the beamformed spectra and imaging data were calibrated using celestial sources (Cassiopeia A, Taurus A) for complex gain, bandpass, and absolute flux calibration. To account for ionospheric refraction effects, we applied refraction corrections based on the quiet Sun location. This correction is essential for accurate source localization at low radio frequencies, where ionospheric effects can significantly distort source positions.

\subsection{Spatiotemporal Evolution in the Dynamic Spectrum} \label{subsec:dynamic_spec}

\begin{figure}[ht!]
    \centering
    \includegraphics[width=0.85\linewidth]{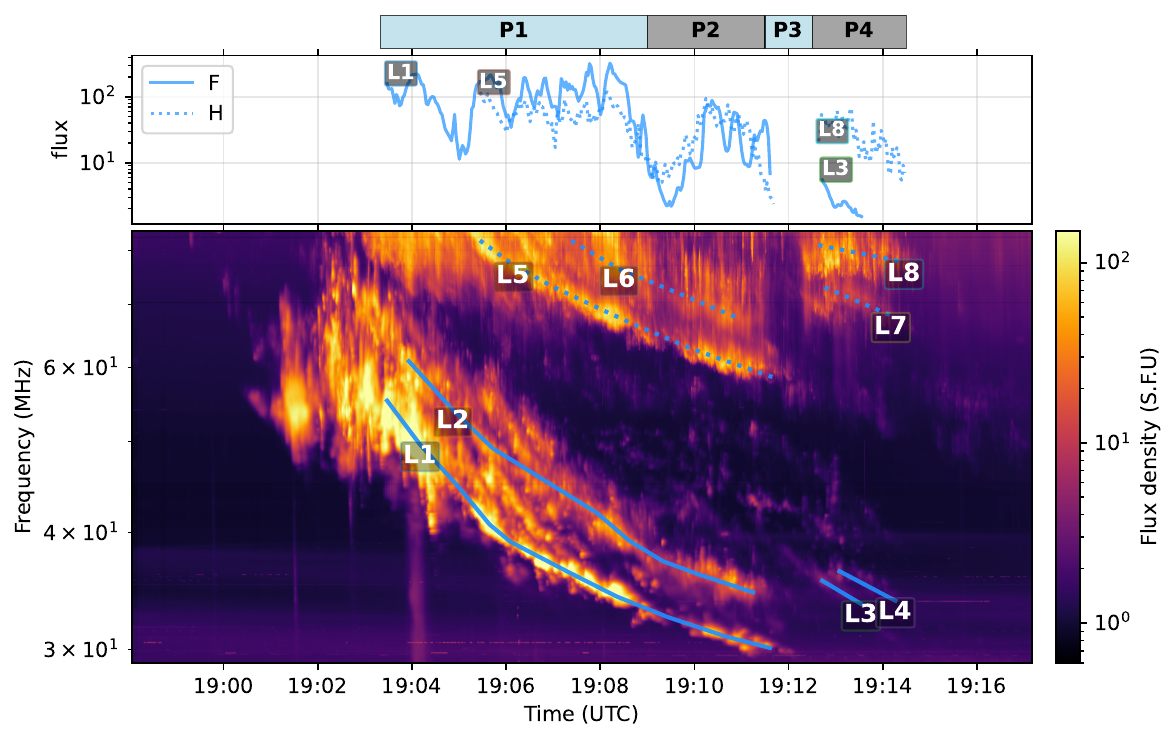}
    \caption{Type II radio burst event observed on 2024 November 18. \textit{Upper panel:} flux density (logarithmic scale) along the fundamental band (solid blue) and harmonic band (dotted light blue), with gray labels marking selected lanes (e.g., L1, L5 in Phase~1; L3, L8 in Phase~4) that correspond to the tracks in the spectrogram. \textit{Lower panel:} dynamic spectrum (frequency vs.\ time) with flux density in solar flux units (sfu; color scale). The burst shows band splitting with eight lanes labeled L1--L8 (L1--L4: fundamental; L5--L8: harmonic), with fundamental tracks shown as solid blue lines and harmonic tracks as dotted light blue lines. Four phases (P1--P4) are marked above the panels. The upper panel highlights the evolution of relative brightness between fundamental and harmonic emission across phases.}
    \label{fig:spec} 
\end{figure}

Figure~\ref{fig:spec} shows the beamformed dynamic spectrum of the type II radio burst together with integrated flux light curves. The event exhibits clear fundamental--harmonic pairs and band splitting features throughout most of its duration and can be divided into four distinct phases, marked as P1–P4 in the figure. \pzred{The upper panel tracks the flux of the fundamental ($F$) and harmonic ($H$) bands, allowing a direct comparison of the harmonic-to-fundamental intensity ratio $I_H/I_F$ as the burst evolves.}

\begin{itemize}
    \item \textbf{Phase 1} (P1) represents the initial bright phase of the burst, with peak flux densities reaching approximately 150~sfu. This phase is characterized by band splitting with complex lane structures. The fundamental emission lanes are significantly stronger than their harmonic counterparts. \pzred{Throughout P1, $I_H/I_F < 1$.}

    \item \textbf{Phase 2} (P2) shows weaker overall emission with less complex lane structures. Notably, the relative intensity between fundamental and harmonic emission changes, with the fundamental becoming comparable to or less dominant than the harmonic. \pzred{Here $I_H/I_F$ approaches unity and occasionally exceeds unity, marking a transition from $F$-dominated to $H$-dominated emission.} 

    \item \textbf{Phase 3} (P3) is marked by a brief disappearance of the type II emission, with significantly reduced radio flux for both the fundamental and harmonic emission. 

    \item \textbf{Phase 4} (P4) shows a resurgence of type II emission with a notable jump to higher frequencies in the drift rate. In this phase, in contrast to Phase 1, the fundamental emission is significantly weaker than its harmonic counterpart. \pzred{The harmonic clearly dominates, with $I_H/I_F > 1$.}
\end{itemize}

\pzred{The systematic evolution from $I_H/I_F < 1$ in P1 to $I_H/I_F \gtrsim 1$ in P2 and P4 shows that the relative strength of the $F$ and $H$ bands is not fixed but changes with time within a single event. }

We identify eight distinct emission lanes in the dynamic spectrum, labeled L1–L8 in Figure~\ref{fig:spec}. Lanes L1–L4 correspond to fundamental emission, while lanes L5–L8 represent their harmonic counterparts. The band splitting is evident in both the fundamental and harmonic components, with each showing two parallel lanes that drift slowly toward lower frequencies as the shock propagates outward.

\subsection{Spatial Evolution of Radio Sources}

\begin{figure}[ht!]
    \centering
    \includegraphics[width=0.85\linewidth]{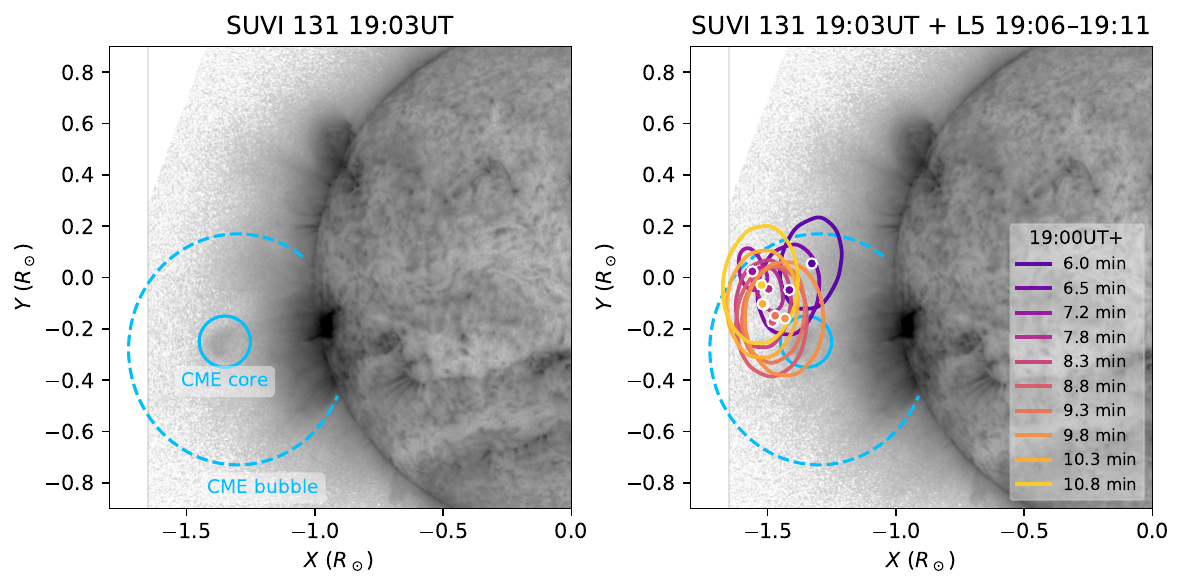}
    \caption{OVRO--LWA radio sources comparing with the SUVI EUV image at 19:03 UT on 2024 November 18. 
    The EUV emission (in blue) presents a CME core and a fainter CME bubble; 
    the radio source markers (colored from purple to red to yellow) show where the type II emission of L5 is located relative to the CME.}
    \label{fig:overplot}
\end{figure}

By comparing the OVRO--LWA radio source positions with the coronal eruption observed in extreme ultraviolet (EUV) images by the GOES Solar Ultraviolet Imager \citep[SUVI;][]{darnel2022suvi}, as shown in Figure~\ref{fig:overplot}, we examine where the radio emission originates relative to the erupting CME structure. The EUV image shows a relatively well-defined CME core surrounded by a fainter CME bubble; the radio sources \pzred{of lane L5 from the harmonic band, in the frequency range of 60--80~MHz} are overlaid to show their location relative to the erupting structure, suggesting that the radio source tends to originate from the flank region of the CME.
\pzred{Harmonic emission is preferably used to localize the radio source because it is less affected by the scattering and refraction from CME and coronal structures \citep{zhang2021parametric}.}

\begin{figure}[tp!]
    \centering
    \includegraphics[width=0.96\linewidth]{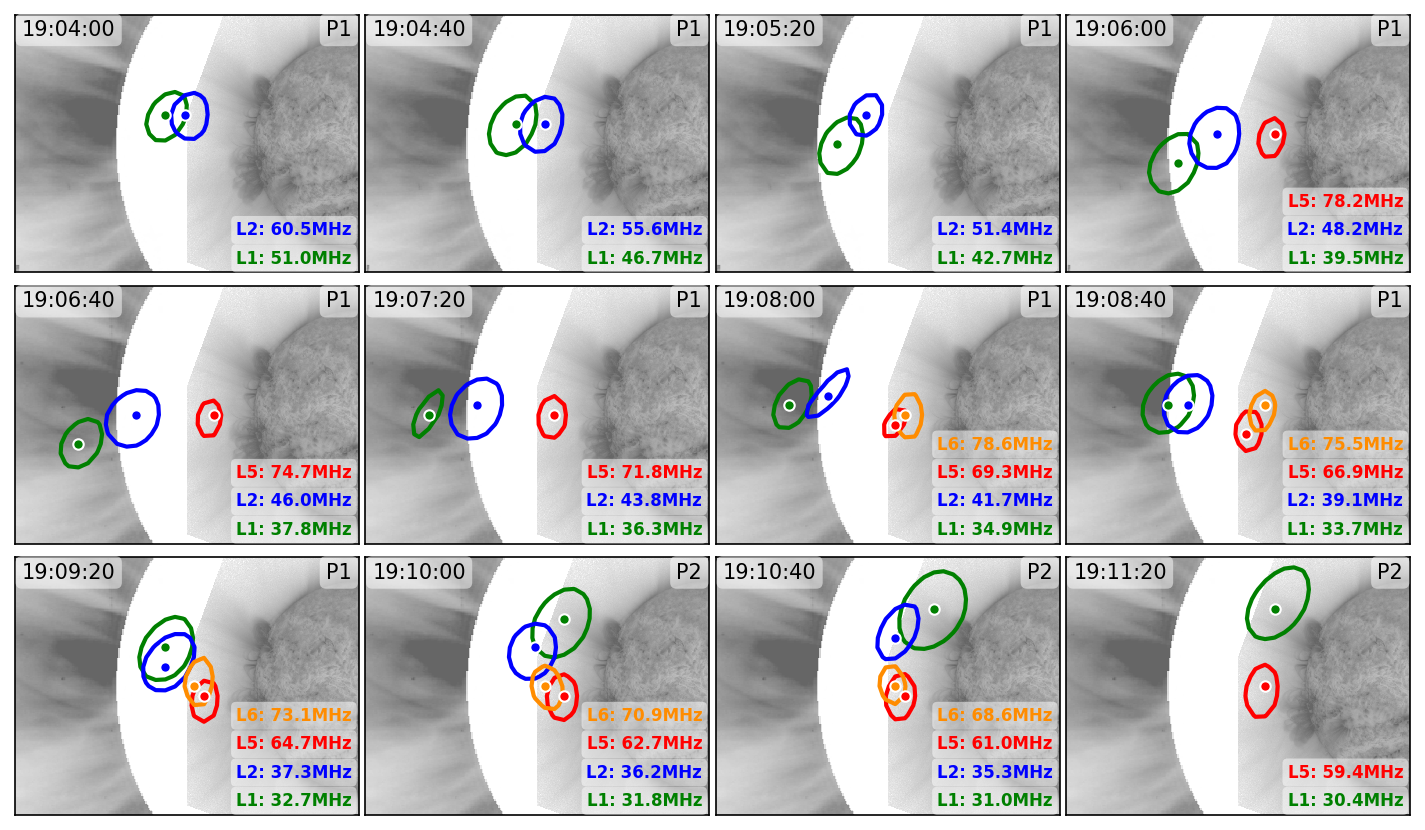}
    \caption{Time evolution of radio source positions during Phases 1 and 2 for lanes L1, L2 (fundamental), and L5, L6 (harmonic). The spatial separation between fundamental and harmonic sources, as well as between split-band lanes, provides constraints on the 3D shock geometry.}
    \label{fig:p1p2}
\end{figure}

Figure~\ref{fig:p1p2} shows the time evolution of the radio source positions for lanes L1, L2, L5, and L6 during Phases 1 and 2. The fundamental and harmonic sources are spatially separated, with the fundamental sources shifting into different directions. 
The spatial separation between the fundamental and harmonic sources, as well as between the split-band lanes shown in Figure~\ref{fig:spec}, provides constraints on the 3D shock geometry and the spatial structure of the emission regions.

\begin{figure}[tp!]
    \centering
    \includegraphics[height=0.32\linewidth]{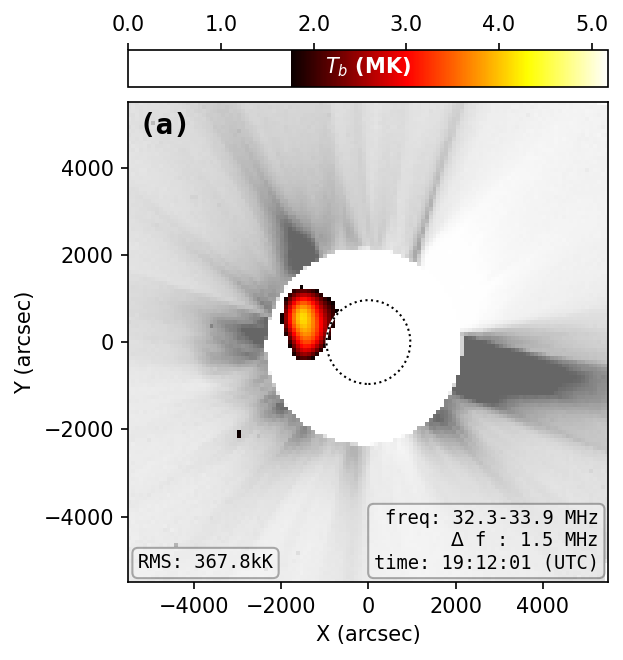}
    \includegraphics[height=0.32\linewidth]{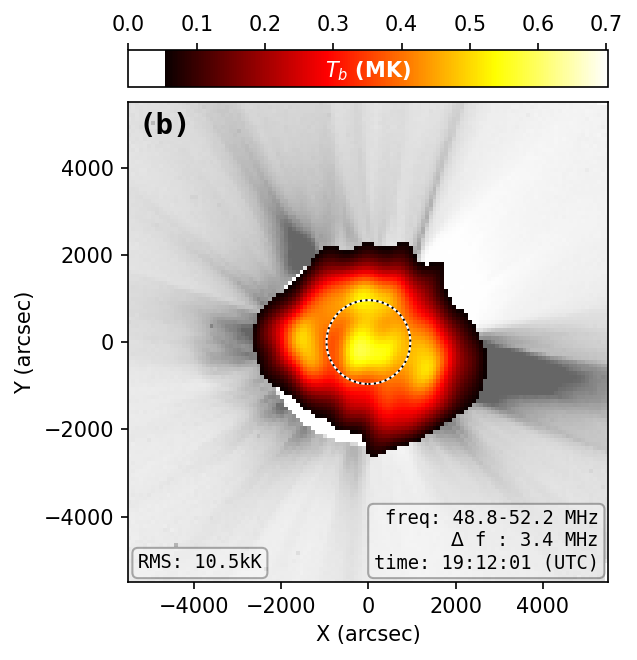}
    \includegraphics[height=0.32\linewidth]{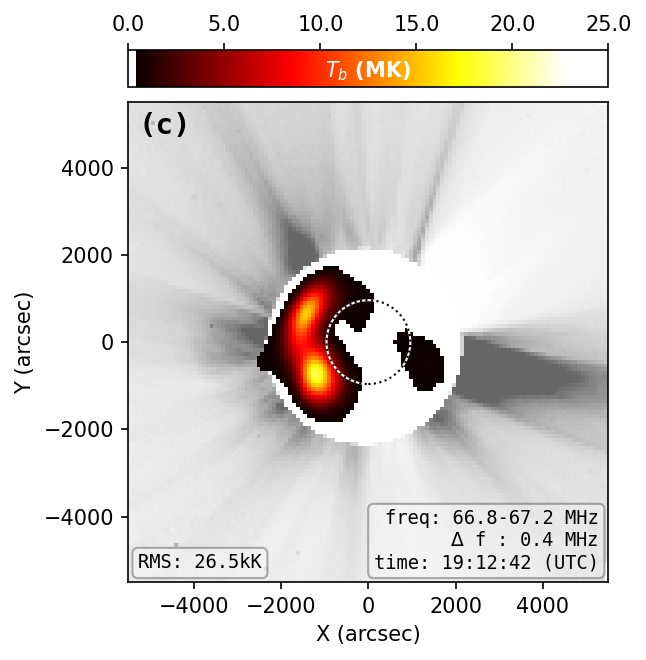}
    \caption{Radio imaging during Phase 3, the period when type II emission temporarily disappears. Left panel: fundamental frequency imaging; middle panel: quiet Sun reference; right panel: harmonic frequency imaging. Note that the color bar range is different in each panel.} 
    \label{fig:p3}
\end{figure}

The radio source variation is shown in Figure~\ref{fig:p3} during the period when the type II emission temporarily disappears from the dynamic spectrum. The panels show the imaging at the fundamental frequency, a quiet Sun reference, and the harmonic frequency, providing context for the burst environment. The brightness temperature is also significantly lower than the source in other phases  \ref{tab:peak_Tb}, where no strong localized radio source is present during this interval. 
This is consistent with the dynamic spectrum and suggests a temporary reduction or interruption in type II radio emission.

\begin{table}[ht!]
    \centering
    \caption{Peak Brightness Temperature}
    \label{tab:peak_Tb}
    \hspace*{-3.5cm}
    \begin{tabular}{lccccccccccc}
    \hline\hline
        Component & L1 & L2 & L5 & L6 & P3 (a) & P3 (b) & P3 (c) & L3 & L4 & L7 & L8 \\
    \hline
        Band & F & F & H & H & F & --- & H & F & F & H & H \\
        $T_{\mathrm{b,peak}}$ (MK)
        & \colorbox{red!29}{3077}
        & \colorbox{red!22}{743}
        & \colorbox{red!24}{1032}
        & \colorbox{red!20}{479}
        & \colorbox{yellow!15}{4.145}
        & \colorbox{blue!9}{0.586}
        & \colorbox{orange!13}{18.987}
        & \colorbox{yellow!17}{10.6}
        & \colorbox{orange!12}{31.0}
        & \colorbox{orange!17}{105}
        & \colorbox{red!18}{283} \\
    \hline
    \end{tabular}
\end{table}

\begin{figure}[ht!]
    \centering
    \includegraphics[width=0.6\linewidth]{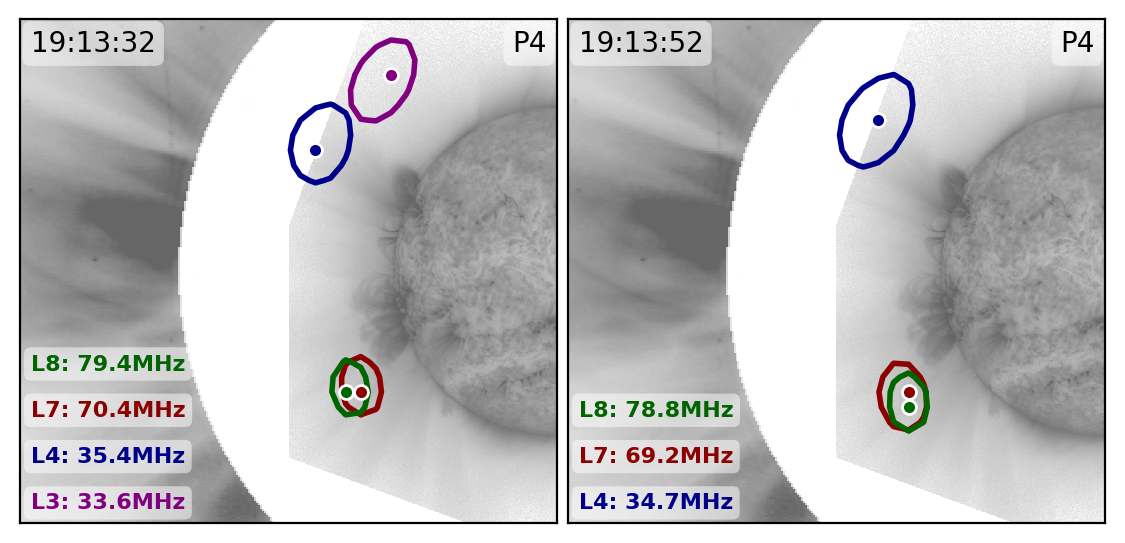}
    \caption{Radio imaging during Phase 4, when type II emission brightens again. A plotting style similar to that in Figure~\ref{fig:p1p2} is used, with L3 and L4 corresponding to the fundamental emission and L7 and L8 corresponding to the harmonic emission.} 
    \label{fig:p4}
\end{figure}

Figure~\ref{fig:p4} displays the imaging results for Phase 4, when the type II emission brightens again. In contrast to Phases 1 and 2 (see Figure~\ref{fig:p1p2}), the radio sources during this phase appear at different projected locations, indicating that the emission region has changed as the eruption evolves. 
This change may reflect the interaction of the CME-driven shock with different coronal structures, leading to modified local shock conditions and particle acceleration efficiency. We also note that the harmonic emission dominates in this phase, as indicated by the dynamic spectrum in Figure~\ref{fig:spec}.

Together, these imaging results in four different phases show that the type II source locations evolve systematically during the early stage of the CME eruption. The emission occurs preferentially away from the CME core. The harmonic part of the split-bands (L5 and L6) is mostly co-located during P1 and P2. There are clear differences between the fundamental and harmonic source locations. This spatial evolution suggests that the radio burst is controlled not only by the global outward propagation of the CME-driven shock, but also by the local shock geometry and coronal environment encountered by different parts of the shock surface.

\section{MHD Simulation} \label{section:simu3}

To interpret the observed evolution of the type II radio sources and characterize the 3D geometry of the CME-driven shock, we employ AWSoM-R\footnote{Available at \url{https://github.com/SWMFsoftware} and \url{https://ccmc.gsfc.nasa.gov/ror/requests/SH/SWMF-AWSoM/swmfawsom_user_registration.php}.} \citep{sokolov2021threaded}, a 3D global MHD model that enables realistic simulations of global steady-state solar wind, CME flux rope generation and propagation, and shock formation \citep[see also][]{sokolov2013magnetohydrodynamic, van2014alfven}. Our simulation is driven by the hourly updated, zero-point corrected photospheric synoptic map from the Global Oscillation Network Group \citep[GONG\footnote{\url{https://nso.edu/data/nisp-data/magnetograms/}},][]{harvey1996global, hill2018global} and incorporates data-driven, force-imbalanced Gibson-Low (GL) flux rope parameters for the CME eruption \citep{gibson1998time, borovikov2017eruptive, jin2017data}. As listed in Table \ref{tab1:param}, this model setup enables us to reproduce the large-scale structure and evolution of the 2024 November 18 CME eruption and its associated shock. 

    \begin{table}[ht!]
    \begin{center}
    \caption{AWSoM-R and CME flux rope parameters used for the solar wind and CME simulations.} \label{tab1:param}
    \setlength\tabcolsep{10pt}{
    \begin{tabular}{llc}
    \hline\hline 
        Model & Parameter & Value \\ 
    \hline 
        AWSoM-R & Poynting flux parameter & $0.3\; \mathrm{MW\; m^{-2}\; T^{-1}}$ \\
        & Correlation length for dissipation & $1.5\times10^5 \; \mathrm{m\; T^{1/2}}$ \\
        & Stochastic heating exponent & $0.21$ \\
        & Stochastic heating amplitude & $0.18$ \\ 
    \hline 
        CME flux rope & Type & GL \\
        & Source region location\tablenotemark{$*$} & $\left(217.50^{\circ},\, -7.36^{\circ}\right)$ \\
        & Major axis orientation & $10.24^{\circ}$ \\
        & Helicity & $+1$ \\
        & Radius & $0.80 \;R_\mathrm{s}$ \\
        & Stretch & $0.60 \;R_\mathrm{s}$ \\
        & Apex height & $1.00 \;R_\mathrm{s}$ \\
        & Magnetic field strength & $8.40 \;\mathrm{G}$ \\
    \hline
    \end{tabular}}
    \end{center}
    \tablenotetext{*}{ This location is given as the Carrington longitude and latitude. }
    \end{table}

Within the AWSoM-R simulation, we identify the CME-driven shock front and systematically characterize the shock properties on the shock surface using a shock--capturing tool \citep{liu2025physics, chen2025evidence}. The resulting output quantities include the shock obliquity angle $\theta_{Bn}$ and the Alfv\'{e}nic Mach number $M_\mathrm{A}$, the density compression ratio, and other parameters relevant to particle acceleration. In particular, quasi-perpendicular shocks ($\theta_{Bn} > 45^\circ$) are generally believed to be more efficient at accelerating electrons than quasi-parallel shocks \citep[e.g.,][]{kong2020quasiperpendicular, hegedus2021tracking}, making $\theta_{Bn}$ a key diagnostic for understanding the spatial distribution of type II radio emission.

\begin{figure}[tp!]
    \centering
    \includegraphics[width=0.99\linewidth]{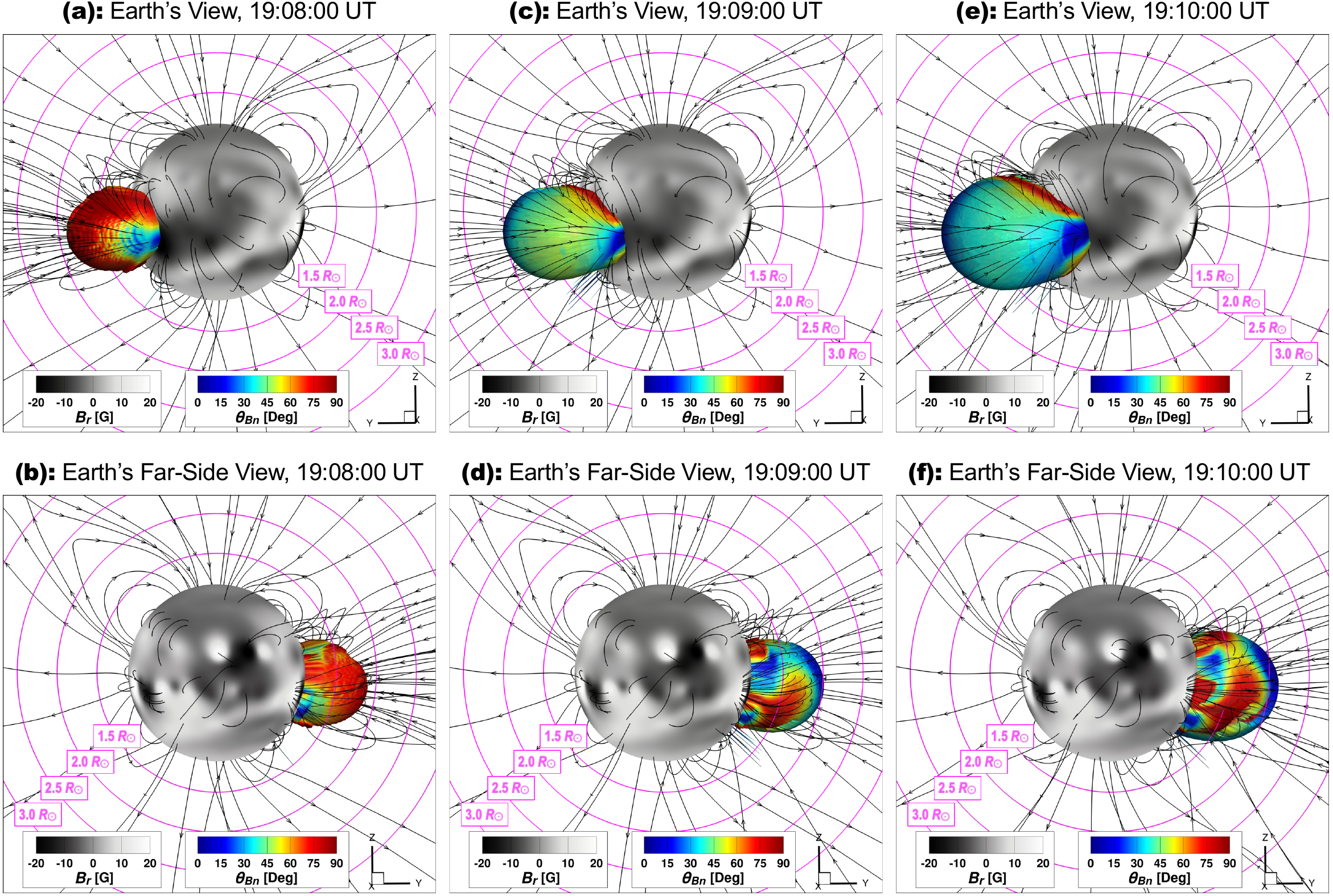}
    \caption{The Sun and the 3D shock front identified by the shock-capturing tool in the AWSoM-R simulation. From left to right, the columns correspond to 19:08:00 UT, 19:09:00 UT, and 19:10:00 UT, representing Phase 1, the Phase 1–2 transition, and Phase 2, respectively. In each column, the upper and lower panels show the Earth-side and far-side views, respectively. Each panel shows the radial magnetic field strength ($B_r$) of the Sun at the lower boundary in the AWSoM-R simulation ($r = 1.1~R_\odot$), overlaid with the captured shock front characterized by the shock obliquity angle ($\theta_{Bn}$). Selected magnetic field lines are shown as black curves with arrows, and magenta concentric circles mark heliocentric distances from $1.5\; R_\odot$ to $3.0\; R_\odot$ in the plane of the sky. Heliographic rotating coordinates are used, with the system rotated such that the negative $X$-axis points toward Earth.}
    \label{fig:mhd1}
\end{figure}

Figure~\ref{fig:mhd1} presents the 3D shock front identified by the shock--capturing tool at three representative times: 19:08:00 UT (Phase 1), 19:09:00 UT (Phase 1–2 transition), and 19:10:00 UT (Phase 2). Each column shows the Earth-facing view (upper panel) and the far-side view (lower panel). The radial magnetic field strength, $B_r$, at the simulation lower boundary ($r = 1.1\; R_\odot$) is displayed on the solar surface, providing context for the ambient coronal field through which the shock propagates. 
The shock front is colored by $\theta_{Bn}$, with red indicating quasi-perpendicular regions ($\theta_{Bn} \approx 90^\circ$) and blue indicating quasi-parallel regions ($\theta_{Bn} \approx 0^\circ$). A large number of magnetic field lines, preferentially sampled near the shock front, are also overplotted as black curves with arrows to further illustrate the large-scale magnetic connectivity and the upstream magnetic-field geometry surrounding the evolving shock front. 

The shock front exhibits a complex, inhomogeneous morphology that reflects the interaction between the expanding CME and the structured coronal magnetic field. 
At 19:08:00 UT, large $\theta_{Bn}$ values are widespread over much of the visible shock surface, while a localized region with smaller $\theta_{Bn}$ appears near the flank of the Earth-facing front. During the transition to 19:09:00 UT and 19:10:00 UT, the Earth-facing side becomes predominantly more parallel, whereas large $\theta_{Bn}$ values are concentrated along the outer rim of the shock and the flanks in Earth's far-side view. 
Further examination of the 3D structure and shock evolution reveals that this rapid and significant change in the shock obliquity angle, rather than a gradual transition \citep[e.g.,][]{manchester2005coronal, liu2025physics}, is caused by the shock propagating from the low corona, where closed-field regions are present, into regions with more open and radially directed magnetic fields. 

\begin{figure}[tp!]
    \centering
    \includegraphics[width=0.99\linewidth]{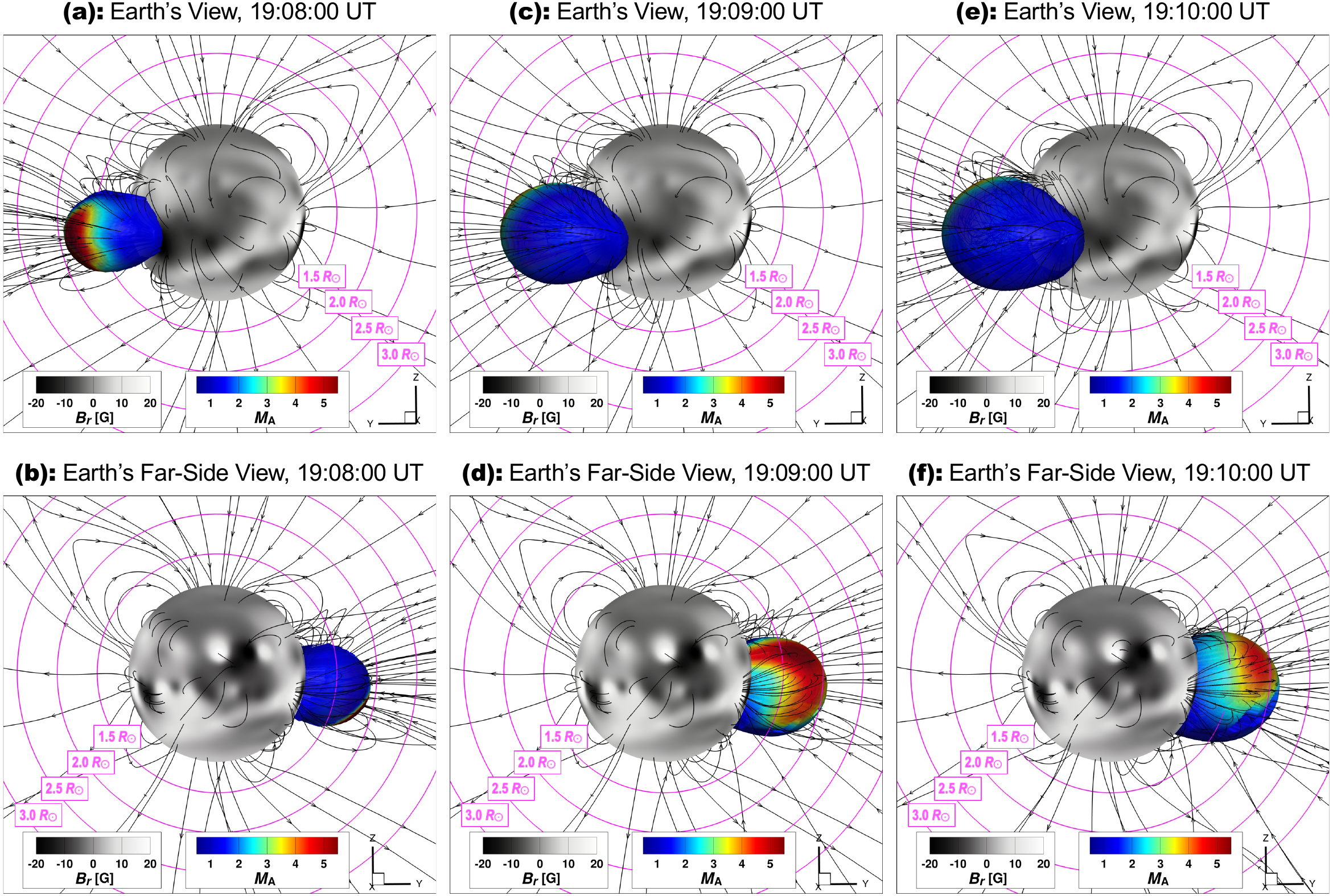}
    \caption{The Sun and the 3D shock front in the AWSoM-R simulation, shown with the same panel layout and plotting style as Figure~\ref{fig:mhd1} except for the shock front colored by the Alfv\'{e}nic Mach number ($M_\mathrm{A}$).}
    \label{fig:mhd2}
\end{figure}

To further examine the shock strength, in Figure~\ref{fig:mhd2}, we plot the same shock front and time sequence, but with the shock surface colored by the Alfv\'{e}nic Mach number, $M_\mathrm{A}$. 
The Mach number is also highly nonuniform across the shock surface. At 19:08:00 UT, enhanced $M_\mathrm{A}$ appears mainly at the nose as can be seen from the Earth-facing shock front, while the opposite side remains relatively weak. At 19:09:00 UT and 19:10:00 UT, the Earth-facing shock surface is dominated by relatively low $M_\mathrm{A}$ values, whereas stronger shocks are found primarily along the far-side flank and outer rim. Hence, regions favorable for efficient particle acceleration and, possibly, producing type II radio bursts are those where the quasi-perpendicular geometry in Figure~\ref{fig:mhd1} overlaps with the enhanced-$M_\mathrm{A}$ regions in Figure~\ref{fig:mhd2}. These regions are located first near the nose and lateral part of the Earth-facing shock in Phase 1 and later mainly along the far-side flank and/or outer rim during the Phase 1–2 transition and Phase 2.

\begin{figure}[tp!]
    \centering
    \includegraphics[width=0.96\linewidth]{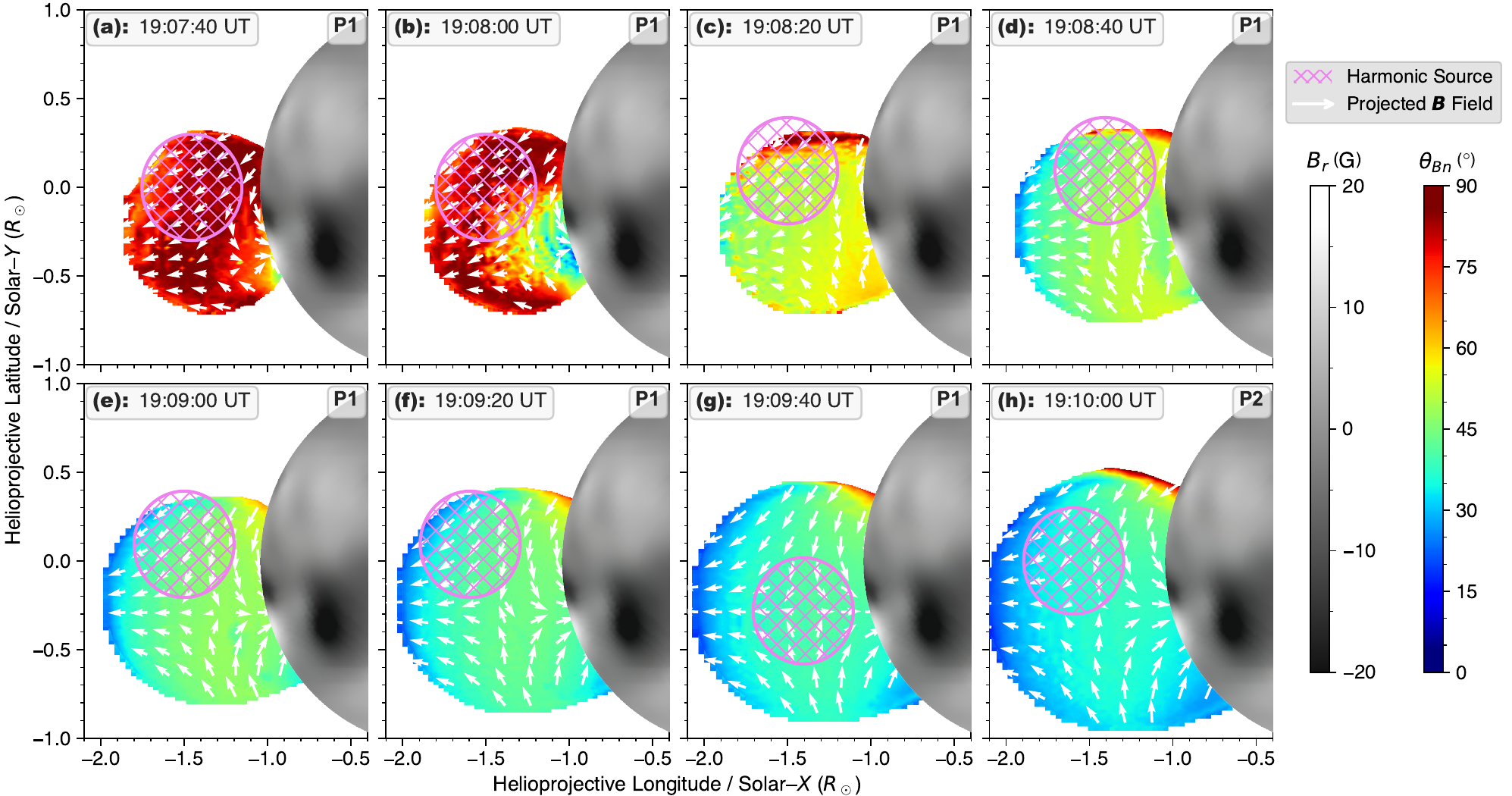}
    \caption{The Sun, the captured shock front, and the 3D magnetic field projected onto the plane of the sky as viewed from Earth. Panels (a)–(h) show snapshots from 19:07:40 to 19:10:00 UT at a cadence of 20 seconds during Phases 1 and 2. Each panel covers the plane-of-sky region of $[-2.1\; R_\odot, \, -0.4\; R_\odot] \times [-1.0\; R_\odot, \, 1.0\; R_\odot]$. 
    The right side of each panel shows part of the $B_r$ distribution on the $r = 1.1\; R_\odot$ sphere, while the left side shows the plane-of-sky shock front colored by $\theta_{Bn}$. The modeled vector magnetic field on the shock surface is projected onto the plane of the sky and shown as white arrows, with the arrow length proportional to the magnetic field strength. The harmonic source region derived from the OVRO--LWA observations is shown in pink with cross markers, with an averaged uncertainty radius of $0.3 \;R_\odot$.}
    \label{fig:mhd3}
\end{figure}

\begin{figure}[tp!]
    \centering
    \includegraphics[width=0.96\linewidth]{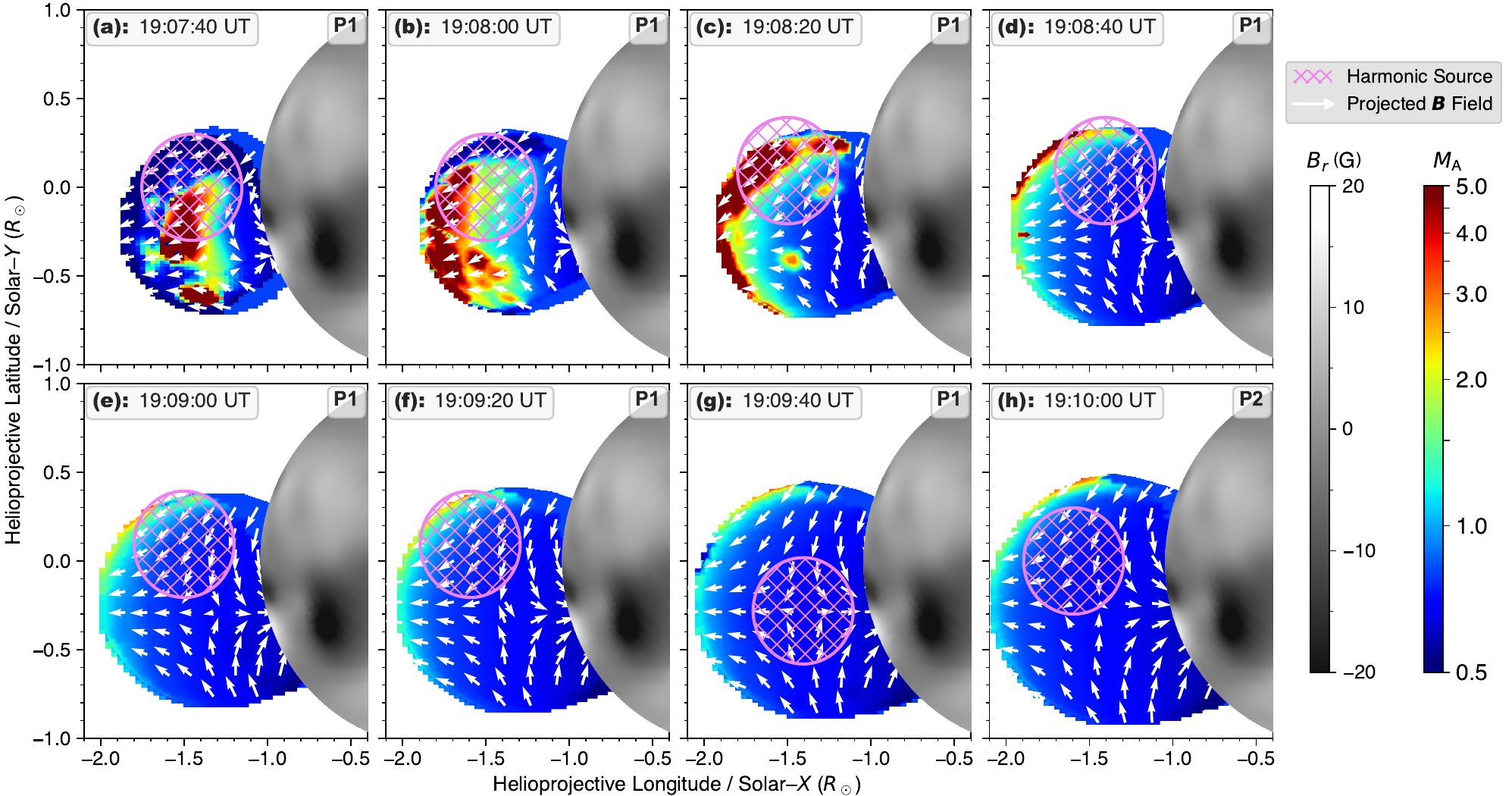}
    \caption{The Sun, the captured shock front, and the 3D magnetic field projected onto the plane of the sky as viewed from Earth, shown with the same panel layout and plotting style as Figure \ref{fig:mhd3} except for the shock front colored by the Alfv\'{e}nic Mach number.}
    \label{fig:mhd4}
\end{figure}

Figures~\ref{fig:mhd3} and~\ref{fig:mhd4} show the shock front and the modeled magnetic field projected onto the plane of the sky as viewed from Earth, enabling a direct comparison with the OVRO--LWA radio imaging as described in Section~\ref{section:obs2}. Panels (a)–(h) show snapshots of the $B_r$ distribution on the $r = 1.1~R_\odot$ sphere (right side of each panel) and projected shock surface with the vector magnetic field (left side of each panel) from 19:07:40 to 19:10:00 UT at a 20~s cadence, matching the imaging cadence of the radio observations. 
The harmonic source region derived from the OVRO--LWA observations is overlaid in pink, with cross markers indicating the source locations and a circle denoting an averaged localization uncertainty radius of $0.3~R_\odot$. \pzred{This value is adopted as half of the average deconvolved source major axis, $\langle B_\mathrm{maj}\rangle \approx 775''$, measured for the harmonic emission in the 60--65~MHz band, which corresponds to $\approx 0.3~R_\odot$ at the heliocentric distance of the sources.}
We focus on the harmonic source since it is less affected by coronal refraction and scattering than the fundamental emission, thus providing a more reliable comparison with the modeled shock location. During the early interval from 19:07:40 UT to 19:08:40 UT (panels (a)–(d)), the harmonic sources lie close to regions of large $\theta_{Bn}$ on the projected shock front and overlap with locally enhanced $M_\mathrm{A}$ regions, supporting the interpretation that type II emission is concentrated in regions with quasi-perpendicular shock geometry and enhanced shock strength. 
After 19:09:00 UT (panels (e)–(h)), however, the Earth-facing side of the projected shock becomes dominated by lower $\theta_{Bn}$ and relatively small $M_\mathrm{A}$ values near the observed source locations. The projected harmonic sources no longer coincide with clearly favorable regions on the Earth-facing shock surface. This is likely because the radio emission originates from the far-side portion of the shock, where the 3D shock in Figures~\ref{fig:mhd1} and \ref{fig:mhd2} shows more quasi-perpendicular geometry and larger $M_\mathrm{A}$ values along the limb and farther flank regions, and is then projected onto the plane of the sky. This projection effect can explain why the radio source remains close to the projected shock surface even when the Earth-facing surface itself does not show locally favorable shock conditions (see also the discussion in Section~\ref{section:sum5}). 

\section{Radio Source and Shock Plasma Condition} \label{section:surface4}

The combination of OVRO--LWA radio imaging and AWSoM-R MHD simulation enables a quantitative comparison between the observed type II source locations and the modeled shock geometry. For type II radio emission to be clearly observed, several conditions must be satisfied: (1) a shock must form with $M_\mathrm{A} > 1$; (2) assuming SDA is the relevant electron acceleration mechanism, the shock geometry favorable for generating suprathermal electrons is typically associated with a quasi-perpendicular geometry, i.e., $\theta_{Bn} \gtrsim 45^\circ$; and (3) the emitted radio waves must be able to propagate to the observer, which can be affected by refraction, scattering, and line-of-sight (LOS) geometry. The alignment of the observed radio sources with regions of locally enhanced $M_\mathrm{A}$ and large $\theta_{Bn}$ on the modeled shock front (Figures~\ref{fig:mhd3} and~\ref{fig:mhd4}) supports the interpretation that type II emission is preferentially generated in regions where the shock normal is nearly perpendicular to the upstream magnetic field.

\subsection{Band Splitting and Shock Plasma Properties} \label{subsec:bandshock4.1}

The band splitting feature observed in the dynamic spectrum (Figure~\ref{fig:spec}) provides additional constraints on the shock plasma conditions. A commonly adopted interpretation attributes the two nearly parallel lanes to plasma emission from the upstream and downstream regions of the shock \citep{Smerd1974, Vrsnak2002}. In this event, the fundamental and harmonic sources for each split-band pair are located at similar positions in the plane of the sky (Figure~\ref{fig:p1p2}), consistent with an upstream--downstream scenario in which both emission regions are located along a similar LOS through the shock front.

The frequency ratio between the upper and lower lanes of the band splitting is related to the density compression ratio across the shock. For plasma emission at either the fundamental or harmonic, the frequency ($f$) is proportional to $n^{1/2}$ with $n$ denoting the plasma number density, so the frequency ratio $f_\mathrm{upper}/f_\mathrm{lower}$ reflects the square root of the density jump. The observed band splitting ratio of about 1.15--1.2 in this event corresponds to a density compression ratio of $\sim$1.3--1.4. Applying the Rankine--Hugoniot relations for a fast MHD shock \citep{Vrsnak2002}, this compression ratio implies an Alfv\'{e}nic Mach number in the range $M_\mathrm{A} \sim 1.5$--2.5, with a dependence on $\theta_{Bn}$ and the plasma $\beta$ \citep[see also][]{zucca2014understanding, maguire2020evolution}. This estimate is consistent with the $M_\mathrm{A}$ values of 1--4 displayed on the shock front in the AWSoM-R simulation (Figure~\ref{fig:mhd4}), with the radio-emitting regions coinciding with portions of the shock where both $\theta_{Bn}$ and $M_\mathrm{A}$ are favorable for efficient particle acceleration.

\subsection{Fundamental--Harmonic Intensity Ratio} \label{subsec:fh_ratio}

We note that the relative brightness between the fundamental and harmonic lanes varies systematically across the four phases (Figure~\ref{fig:spec}). During Phase~1, the fundamental lanes (L1--L4) dominate over their harmonic counterparts (L5–L8), whereas in Phase~2 the harmonic becomes comparable to or stronger than the fundamental, and by Phase~4 the harmonic emission clearly dominates. Because the F and H bands are produced by the same shock-accelerated electron population but propagate through the corona at different frequencies and with different angular patterns, the observed F/H ratio can reflect both intrinsic emission properties and LOS propagation effects.

\begin{figure}[ht!]
    \centering
    \includegraphics[width=0.8\linewidth]{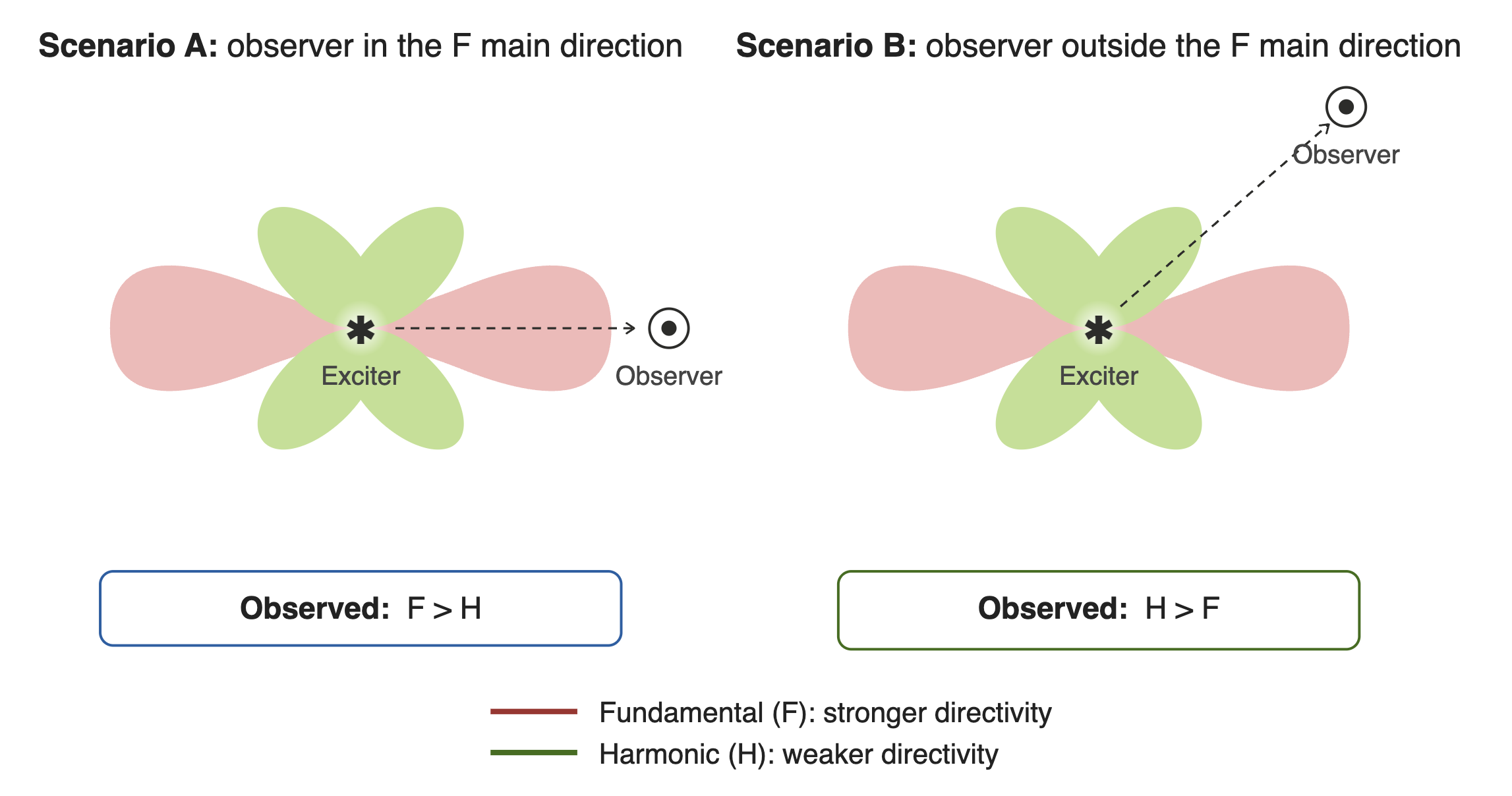}
    \caption{Schematic illustrating how differences in emission directivity can affect the observed fundamental-to-harmonic brightness ratio. In Scenario A (left), the observer lies near the fundamental (F) main emission direction, yielding F$\;>\;$H. In Scenario B (right), the observer is outside the main direction of the fundamental; the harmonic (H), with weaker directivity, is more readily observed and can dominate, yielding H$\;>\;$F.}
    \label{fig:beampattern}
\end{figure}

The \pzred{directivity and propagation effects} are important in this event. 
First, fundamental plasma emission is expected to be more strongly directional than harmonic emission \citep{reid2014typeIII}. The fundamental is generated close to the local plasma frequency and is therefore more sensitive to the local refractive index and scattering environment, which can confine the escaping radiation to a narrower angular range \citep{robinson2000typeIII}. 
In contrast, the harmonic is emitted at a higher frequency, about twice the plasma frequency, and typically has a broader emission pattern. In this scenario, as illustrated in Figure~\ref{fig:beampattern}, an observer located near the main emission direction of the fundamental would measure F$\;>\;$H, while an observer outside this direction receives strongly suppressed fundamental emission but appreciable harmonic emission that is less directional, resulting in H$\;>\;$F.
Second, propagation through the structured corona can suppress the fundamental emission along some viewing geometries. Compared with the harmonic, the fundamental is more susceptible to refraction away from the LOS and to effective occultation when the ray path approaches dense coronal structures or passes behind the limb. As the modeled shock and radio-emitting region evolve from an Earth-facing configuration (early Phase~1; panels (a)–(d) of Figures~\ref{fig:mhd3} and~\ref{fig:mhd4}) toward a geometry in which the emission is more likely to originate from the far-side shock surface (late Phase~1 to Phase~2; panels (e)–(h)), the fundamental becomes increasingly disadvantaged relative to the harmonic. This trend can explain the transition from fundamental-dominated emission in Phase~1 to harmonic-dominated emission in Phases~2 and~4 without requiring a change in the underlying shock existence; rather, as indicated by the simulation results, the changing 3D source geometry and propagation conditions regulate which band can effectively reach the observer.

\subsection{Fundamental--Harmonic Source Offset and Magnetic Field} \label{subsec:fh_offset}

The spatial offset between the fundamental and harmonic radio sources primarily arises from propagation effects that affect the fundamental emission much more strongly than the harmonic. \pzred{The harmonic emission, produced at about twice the local plasma frequency, is negligibly affected by propagation effects (scattering), and its observed position closely reflects its original source location. In contrast, the fundamental emission, with its lower frequency, is subject to stronger scattering in the corona. Notably, anisotropic scattering in a magnetized turbulent plasma can shift the apparent fundamental source position preferentially along the direction of the local magnetic field \citep{clarkson2025mag}. As a result, the observed offset direction between the fundamental and harmonic sources is governed by the direction of the local magnetic field in the emission region, making this offset a sensitive diagnostic of the projected transverse magnetic field on the shock surface.}

\begin{figure}[ht!]
    \centering
    \includegraphics[width=0.8\linewidth]{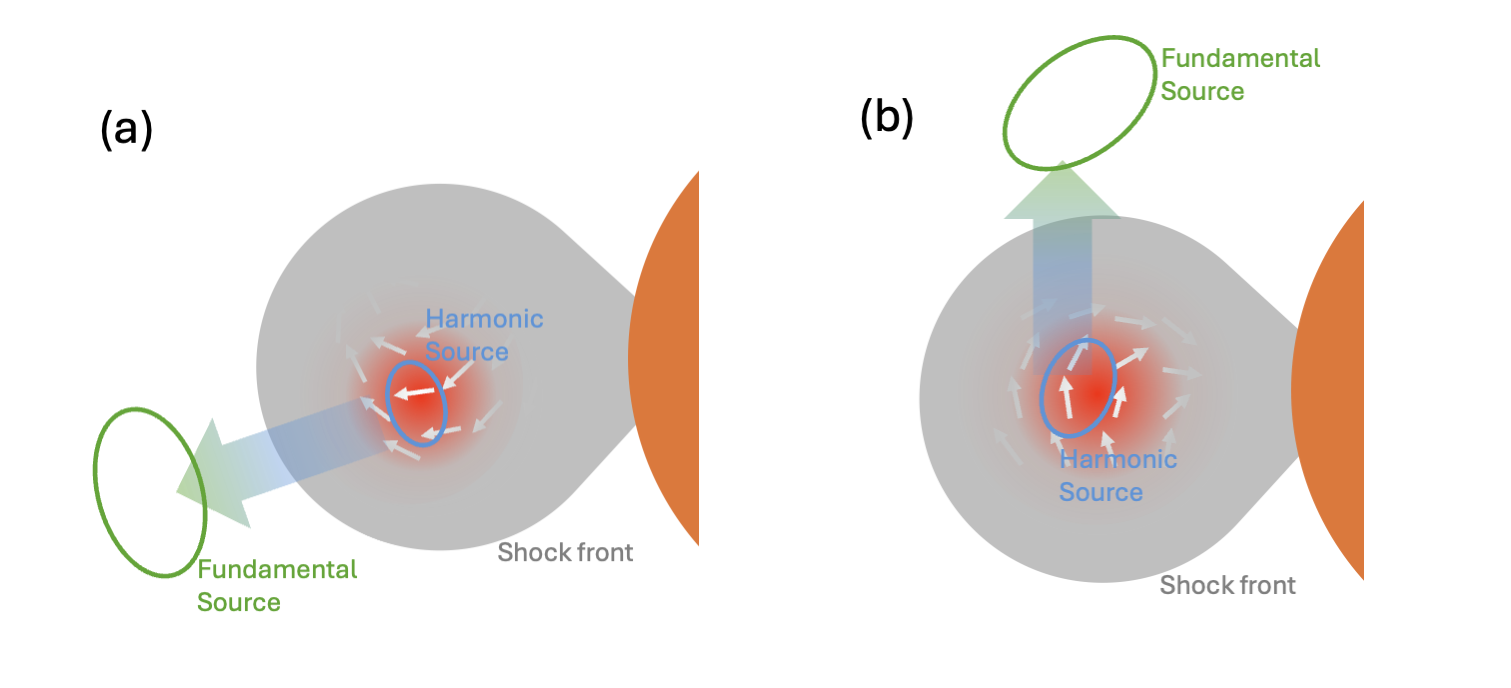}
    \caption{Schematic illustrating how anisotropic scattering of radio waves in the turbulent coronal plasma shifts the apparent radio source position. The offset between the fundamental and harmonic sources reflects the magnetic field direction in the emission region, as the scattering is preferentially enhanced along the field.}
    \label{fig:scattering}
\end{figure}

Figure~\ref{fig:scattering} illustrates how scattering in a magnetized plasma shifts the apparent source position. To compare this interpretation with the MHD simulation, in Figures~\ref{fig:mhd3} and~\ref{fig:mhd4}, the white arrows depict the modeled magnetic field on the shock surface projected onto the plane of the sky. At the beginning of Phase~1, the vectors within the harmonic source region are relatively well aligned and point in a consistent direction, matching the relatively coherent offset between fundamental and harmonic sources in the radio imaging. In later Phase~1 and early Phase~2 (panels (g) and (h)), the harmonic source region overlaps with a region where the projected field vectors are more complex and less unidirectional. This evolution is quantitatively consistent with the trend in the observed radio source offset: at early times the shift between model and observation is more radial-like, while at later times it rotates toward other directions as the source moves into a region with a more structured magnetic field. Overall, the observed offset interpretation and the MHD-derived field orientation are consistent, supporting a connection between radio source morphology and inhomogeneous shock-surface magnetic geometry.

\section{Summary and Conclusions} \label{section:sum5}

We have presented OVRO--LWA imaging spectroscopy of a multi-lane type II radio burst with split-band features on 2024 November 18, in which both fundamental and harmonic emission are clearly visible and spatially resolved. Commensurate with the radio observations, we also conduct a 3D global MHD simulation of the associated CME-driven shock using AWSoM-R to interpret the observed radio morphology in terms of the 3D shock geometry and propagation effects. We summarize the main findings and their implications below.


Imaging results show that the type II radio sources of the higher- and lower-frequency bands within each fundamental--harmonic pair are located at nearly the same position in the plane of the sky (Figure~\ref{fig:p1p2}). This spatial coincidence favors the shock upstream--downstream interpretation \citep{Smerd1974, Vrsnak2002}: the two lanes originate from plasma emission in the upstream and downstream regions of the shock, which are located along the same LOS but emit at slightly different frequencies. The frequency ratio of the split bands yields a density compression ratio of $\sim$1.3--1.4 and an Alfv\'{e}nic Mach number of $M_A\approx 1.5$--2.5, consistent with the shock properties in the MHD simulation (Section~\ref{subsec:bandshock4.1}).


A notable feature of this event is the evolution of the relative brightness between fundamental and harmonic emission across different phases. In Phase~1, the fundamental is significantly stronger than the harmonic; in Phase~2, they become comparable, or the harmonic gradually becomes dominant; and in Phase~4, the harmonic clearly dominates. \pzred{Two explanations may account for this trend.
First, the fundamental is expected to have stronger directivity and to be more sensitive to coronal propagation effects than the harmonic \citep{reid2014typeIII, robinson2000typeIII}; see Section~\ref{subsec:fh_ratio} and Figure~\ref{fig:beampattern}. In the early phases, the fundamental emission is more favorably directed toward Earth, so the observer lies nearer the main escape direction of the fundamental and measures F$\;>\;$H. In later phases, the fundamental beam may be increasingly steered away from the observer, while the less directional harmonic remains detectable, yielding H$\;>\;$F even if both bands arise from a similar shock region.
Second, the AWSoM-R simulation indicates an evolving Earth-facing versus far-side source configuration (Section~\ref{section:simu3}). When the radio-emitting region lies on the Earth-facing side of the shock, both the fundamental and harmonic can propagate toward the observer along the LOS with relatively little obstruction. When the emission is more likely to originate from the limb or far-side shock surface, the fundamental is preferentially suppressed by refraction, absorption, and LOS blocking, whereas the harmonic, at roughly twice the plasma frequency, can still reach the observer. The transition from fundamental-dominated to harmonic-dominated emission is therefore also consistent with the changing 3D shock geometry inferred from the MHD simulation.}

\pzred{A further consideration is whether radio emission from a far-side shock can remain visible to the observer. In an idealized CME--shock geometry, the LOS to a far-side source would need to penetrate the CME and its surrounding shocked plasma. However, the CME interior is not necessarily uniformly denser than the outer shell and may contain cavities or other complex density structures that do not fully block radio propagation. In addition, as shown in panels (e)--(h) of Figures~\ref{fig:mhd3} and~\ref{fig:mhd4}, the Earth-facing portion of the projected shock in the later phase does not exhibit clearly favorable quasi-perpendicular geometry or enhanced $M_\mathrm{A}$, whereas the 3D global MHD modeling results still show suitable conditions along the far-side flank and limb. This supports the interpretation that the observed late-phase radio sources are associated with a far-side emitting region that is projected onto the plane of the sky rather than with a well-developed shock on the Earth-facing surface.}

The spatial offset between the fundamental and harmonic radio sources provides an additional diagnostic of the magnetic field orientation in the emission region. As demonstrated by \citet{clarkson2025mag}, anisotropic scattering of radio waves in the turbulent coronal plasma shifts the apparent source position along the magnetic field direction. 
In this event, the fundamental–harmonic offset direction is consistent with the magnetic field orientation obtained from the AWSoM-R simulation (Figures~\ref{fig:mhd3} and~\ref{fig:mhd4}): the white arrows show the projected magnetic field on the shock, and the evolution from a well-aligned field in early Phase~1 to a more complex field structure in later Phase~1 and early Phase~2 matches the evolution of the observed source offset from a more radial-like shift to a rotated direction (Figure~\ref{fig:p1p2}). The fundamental--harmonic offset therefore provides an observational diagnostic of the shock-surface magnetic field that is both qualitatively and quantitatively consistent with the MHD results.


Together, the unprecedented OVRO--LWA radio spectral imaging and the AWSoM-R MHD simulation establish a direct connection between the complex spatial, temporal, and spectral evolution of the multi-lane type II radio source locations and the modeled coronal shock magnetic geometry, demonstrating the value of combining low-frequency radio imaging with 3D global MHD modeling for diagnosing CME-driven shocks. Studies of this kind will pave the way for a more comprehensive understanding of their role in the associated particle acceleration and transport processes.

%
\facilities{OVRO--LWA}

\software{astropy, casatools, AWSoM-R}




\begin{acknowledgments}
P. Z. and B. C. are supported by NASA LWS grant 80NSSC24K1116 awarded to the New Jersey Institute of Technology (NJIT). W. L. and W. M. are supported by the NASA LWS Strategic Capability grant No.~80NSSC22K0892 (SCEPTER) and NASA SWxC grant No.~80NSSC23M0191 (CLEAR). Computational resources supporting this work are provided by the NASA High-End Computing (HEC) Program through the NASA Advanced Supercomputing (NAS) Division at the Ames Research Center\footnote{\url{https://www.nas.nasa.gov/hecc/}} and by the high-performance computing support from the Texas Advanced Computing Center (TACC) Frontera\footnote{\url{https://tacc.utexas.edu/}} at the University of Texas at Austin \citep{stanzione2020frontera}. The OVRO--LWA expansion project was supported by NSF under grant AST-1828784. OVRO--LWA solar operations are supported by NSF grant AGS-2436999 to NJIT.
\end{acknowledgments}

\bibliography{cite}{}

@article{NelsonMelrose1985,
  author  = {Nelson, G. J. and Melrose, D. B.},
  title   = {Type II bursts},
  journal = {Solar Radiophysics: Studies of Emission from the Sun at Metre Wavelengths},
  year    = {1985},
  editor  = {McLean, D. J. and Labrum, N. R.},
  pages   = {333--359},
  publisher = {Cambridge Univ. Press}
}

@ARTICLE{zhang2024AAshock,
       author = {{Zhang}, Peijin and {Morosan}, Diana E. and {Zucca}, Pietro and {Normo}, Sanna and {Dabrowski}, Bartosz and {Krankowski}, Andrzej and {Vocks}, Christian},
        title = "{Imaging spectroscopy of a spectral bump in a type II radio burst}",
      journal = {\aap},
         year = 2024,
        month = apr,
       volume = {684},
          eid = {L22},
        pages = {L22},
          doi = {10.1051/0004-6361/202449365},
archivePrefix = {arXiv},
       eprint = {2403.19451},
 primaryClass = {astro-ph.SR},
       adsurl = {https://ui.adsabs.harvard.edu/abs/2024A&A...684L..22Z}
}

@ARTICLE{Normo2025AAshock,
       author = {{Normo}, S. and {Morosan}, D.~E. and {Zhang}, P. and {Zucca}, P. and {Vainio}, R.},
        title = "{Imaging and spectropolarimetric observations of a band-split type II solar radio burst with LOFAR}",
      journal = {\aap},
         year = 2025,
        month = jun,
       volume = {698},
          eid = {A175},
        pages = {A175},
          doi = {10.1051/0004-6361/202553702},
archivePrefix = {arXiv},
       eprint = {2505.10243},
 primaryClass = {astro-ph.SR},
       adsurl = {https://ui.adsabs.harvard.edu/abs/2025A&A...698A.175N}
}

@ARTICLE{zucca2025AAmultilane,
       author = {{Zucca}, P. and {Zhang}, P. and {Kozarev}, K. and {Nedal}, M. and {Dey}, S. and {Mancini}, M. and {Kumari}, A. and {Morosan}, D.~E. and {Dabrowski}, B. and {Gallagher}, P.~T. and {Krankowski}, A. and {Vocks}, C.},
        title = "{Source location and evolution of a multilane type II radio burst}",
      journal = {\aap},
         year = 2025,
        month = nov,
       volume = {703},
          eid = {A271},
        pages = {A271},
          doi = {10.1051/0004-6361/202554348},
archivePrefix = {arXiv},
       eprint = {2505.12932},
 primaryClass = {astro-ph.SR},
       adsurl = {https://ui.adsabs.harvard.edu/abs/2025A&A...703A.271Z}
}

@ARTICLE{morosan2025AAshocktypeII,
       author = {{Morosan}, D.~E. and {Jebaraj}, I.~C. and {Zhang}, P. and {Zucca}, P. and {Dabrowski}, B. and {Gallagher}, P.~T. and {Krankowski}, A. and {Vocks}, C. and {Vainio}, R.},
        title = "{Resolving spatial and temporal shock structures using LOFAR observations of type II radio bursts}",
      journal = {\aap},
         year = 2025,
        month = mar,
       volume = {695},
          eid = {A70},
        pages = {A70},
          doi = {10.1051/0004-6361/202452775},
archivePrefix = {arXiv},
       eprint = {2502.16934},
 primaryClass = {astro-ph.SR},
       adsurl = {https://ui.adsabs.harvard.edu/abs/2025A&A...695A..70M}
}

@ARTICLE{zhang2021parametric,
       author = {{Zhang}, PeiJin and {Wang}, ChuanBing and {Kontar}, Eduard P.},
        title = "{Parametric Simulation Studies on the Wave Propagation of Solar Radio Emission: The Source Size, Duration, and Position}",
      journal = {\apj},
         year = 2021,
        month = mar,
       volume = {909},
       number = {2},
          eid = {195},
        pages = {195},
          doi = {10.3847/1538-4357/abd8c5},
archivePrefix = {arXiv},
       eprint = {2101.00911},
 primaryClass = {astro-ph.SR},
       adsurl = {https://ui.adsabs.harvard.edu/abs/2021ApJ...909..195Z}
}

@ARTICLE{clarkson2025mag,
       author = {{Clarkson}, Daniel L. and {Kontar}, Eduard P.},
        title = "{Magnetic Field Geometry and Anisotropic Scattering Effects on Solar Radio Burst Observations}",
      journal = {\apj},
         year = 2025,
        month = jan,
       volume = {978},
       number = {1},
          eid = {73},
        pages = {73},
          doi = {10.3847/1538-4357/ad969c},
archivePrefix = {arXiv},
       eprint = {2411.19630},
 primaryClass = {astro-ph.SR},
       adsurl = {https://ui.adsabs.harvard.edu/abs/2025ApJ...978...73C}
}

@ARTICLE{chen2025ApJlocmag,
       author = {{Chen}, Xingyao and {Chen}, Bin and {Yu}, Sijie and {Mondal}, Surajit and {Stiefel}, Muriel Zo{\"e} and {Zhang}, Peijin and {Gary}, Dale E. and {Krucker}, S{\"a}m and {Anderson}, Marin M. and {Bowman}, Judd D. and {Byrne}, Ruby and {Catha}, Morgan and {Chhabra}, Sherry and {D'Addario}, Larry and {Davis}, Ivey and {Dowell}, Jayce and {Hallinan}, Gregg and {Harnach}, Charlie and {Hellbourg}, Greg and {Hickish}, Jack and {Hobbs}, Rick and {Hodge}, David and {Hodges}, Mark and {Huang}, Yuping and {Isella}, Andrea and {Jacobs}, Daniel C. and {Kemby}, Ghislain and {Klinefelter}, John T. and {Kolopanis}, Matthew and {Kosogorov}, Nikita and {Lamb}, James and {Law}, Casey J. and {Mahesh}, Nivedita and {O'Donnell}, Brian and {Plant}, Kathryn and {Posner}, Corey and {Powell}, Travis and {Prayag}, Vinand and {Rizo}, Andres and {Romero-Wolf}, Andrew and {Shi}, Jun and {Taylor}, Greg and {Trim}, Jordan and {Virgin}, Mike and {Vydula}, Akshatha and {Weinreb}, Sandy and {Woody}, David},
        title = "{Measuring the Magnetic Field of a Coronal Mass Ejection from the Low to Middle Corona}",
      journal = {\apjl},
         year = 2025,
        month = sep,
       volume = {990},
       number = {2},
          eid = {L50},
        pages = {L50},
          doi = {10.3847/2041-8213/adfa71},
archivePrefix = {arXiv},
       eprint = {2508.08970},
 primaryClass = {astro-ph.SR},
       adsurl = {https://ui.adsabs.harvard.edu/abs/2025ApJ...990L..50C}
}

@ARTICLE{jha2025relativeFH,
       author = {{Jha}, Rishikesh G. and {Raja}, K. Sasikumar and {Ramesh}, R. and {Kathiravan}, C. and {Monstein}, Christian},
        title = "{Relative Strengths of Fundamental and Harmonic Emissions of Solar Radio Type II Bursts}",
      journal = {\solphys},
         year = 2025,
        month = nov,
       volume = {300},
       number = {12},
          eid = {168},
        pages = {168},
          doi = {10.1007/s11207-025-02589-8},
archivePrefix = {arXiv},
       eprint = {2511.15011},
 primaryClass = {astro-ph.SR},
       adsurl = {https://ui.adsabs.harvard.edu/abs/2025SoPh..300..168J}
}

@article{Smerd1974,
  author  = {Smerd, S. F. and Sheridan, K. V. and Stewart, R. T.},
  title   = {Split-band structure in type II radio bursts},
  journal = {Astrophysics and Space Science},
  year    = {1974},
  volume  = {27},
  pages   = {243--259},
  doi     = {10.1007/BF00645163}
}

@article{Vrsnak2002,
  author  = {Vršnak, B. and Magdalenić, J. and Aurass, H.},
  title   = {Band-splitting of coronal and interplanetary type II bursts. II. Coronal magnetic field and Alfvén velocity},
  journal = {\aap},
  year    = {2002},
  volume  = {396},
  pages   = {673--682},
  doi     = {10.1051/0004-6361:20021300}
}

@article{Kontar2017,
  author  = {Kontar, E. P. and Yu, S. and Kuznetsov, A. A. and Emslie, A. G. and Fleishman, G. D. and Melnik, V. N. and Sharykin, I. N.},
  title   = {Imaging spectroscopy of solar radio burst fine structures},
  journal = {Nature Communications},
  year    = {2017},
  volume  = {8},
  eid     = {1515},
  doi     = {10.1038/s41467-017-01597-8}
}

@article{Morosan2019,
  author  = {Morosan, D. E. and Gallagher, P. T. and Zucca, P. and Fallows, R. and Carley, E. P.},
  title   = {LOFAR imaging of type II solar radio bursts},
  journal = {\aap},
  year    = {2019},
  volume  = {623},
  eid     = {A63},
  doi     = {10.1051/0004-6361/201834078}
}

@article{armstrong1985shock,
  title={Shock Drift Acceleration},
  author={Armstrong, Thomas P and Pesses, Mark E and Decker, Robert B},
  journal={Collisionless Shocks in the Heliosphere: Reviews of Current Research},
  volume={35},
  pages={271--285},
  year={1985},
  doi={10.1029/GM035p0271},
  publisher={Wiley Online Library}
}

@ARTICLE{holman1983sda,
       author = {{Holman}, Gordon D. and {Pesses}, M. E.},
        title = "{Solar Type II Radio Emission and the Shock Drift Acceleration of Electrons}",
      journal = {\apj},
         year = 1983,
        month = apr,
       volume = {267},
        pages = {837},
          doi = {10.1086/160918},
       adsurl = {https://ui.adsabs.harvard.edu/abs/1983ApJ...267..837H}
}

@INCOLLECTION{mann1994quasiparallel,
       author = {{Mann}, G.},
        title = "{Radio Emission from Quasi-Parallel Shock Waves in the Corona}",
   booktitle = {Fragmented Energy Release in Sun and Stars},
       editor = {{van den Oord}, G. H. J.},
      publisher = {Springer},
       address = {Dordrecht},
         year = 1994,
       pages = {199--203},
          doi = {10.1007/978-94-011-1014-3\_27},
       adsurl = {https://ui.adsabs.harvard.edu/abs/1994ApJS...90..577M}
}

@ARTICLE{robinson2000typeIII,
       author = {{Robinson}, P. A. and {Cairns}, I. H.},
        title = "{Theory of Type {III} And Type {II} Solar Radio Emissions}",
      journal = {Geophysical Monograph Series},
         year = 2000,
       volume = {119},
        pages = {37},
          doi = {10.1029/GM119p0037},
       adsurl = {https://ui.adsabs.harvard.edu/abs/2000GMS...119...37R}
}

@ARTICLE{reid2014typeIII,
       author = {{Reid}, Hamish Andrew Sinclair and {Ratcliffe}, Heather},
        title = "{A Review of Solar Type {III} Radio Bursts}",
      journal = {Research in Astronomy and Astrophysics},
         year = 2014,
       volume = {14},
       number = {7},
        pages = {773--804},
          doi = {10.1088/1674-4527/14/7/003},
       adsurl = {https://ui.adsabs.harvard.edu/abs/2014RAA....14..773R}
}

@article{borovikov2017eruptive,
  title={{Eruptive event generator based on the Gibson-Low magnetic configuration}},
  author={Borovikov, Dmitry and Sokolov, Igor V and Manchester, Ward and Jin, Meng and Gombosi, Tamas I},
  journal={Journal of Geophysical Research: Space Physics},
  volume={122},
  number={8},
  pages={7979--7984},
  year={2017},
  doi={10.1002/2017JA024304},
  publisher={Wiley Online Library}
}

@article{chen2011coronal,
  title={Coronal mass ejections: models and their observational basis},
  author={Chen, PF},
  journal={Living Reviews in Solar Physics},
  volume={8},
  number={1},
  pages={1--92},
  year={2011},
  doi={10.12942/lrsp-2011-1},
  publisher={Springer}
}

@article{chen2025evidence,
  title={{Evidence of Time-dependent Diffusive Shock Acceleration in the 2022 September 5 Solar Energetic Particle Event}},
  author={Chen, Xiaohang and Zhao, Lulu and Giacalone, Joe and Sachdeva, Nishtha and Sokolov, Igor V and Toth, Gabor and Cohen, Christina MS and Lario, David and Guo, Fan and Kouloumvakos, Athanasios and others},
  journal={\apj},
  volume={994},
  number={2},
  pages={242},
  year={2025},
  doi={10.3847/1538-4357/ae1227},
  publisher={IOP Publishing}
}

@article{corti2026advancing,
  title={Advancing Heliophysics and Space Weather Modeling through Open Science},
  author={Corti, Claudio and Kuznetsova, Maria M and Reiss, Martin and Yue, Jia and Karpen, Judith T and Arge, Charles Nickolos and Bacchini, Fabio and Bard, Christopher and Bruinsma, Sean L and Caplan, Ronald M and others},
  journal={ESS Open Archive},
  doi={10.22541/essoar.176824639.92354528/v1},
  year={2026}
}

@article{gibson1998time,
  title={{A Time-dependent Three-dimensional Magnetohydrodynamic Model of the Coronal Mass Ejection}},
  author={Gibson, Sarah E and Low, BC},
  journal={\apj},
  volume={493},
  number={1},
  pages={460},
  year={1998},
  doi={10.1086/305107},
  publisher={IOP Publishing}
}

@article{gopalswamy2009soho,
  title={The SOHO/LASCO CME Catalog},
  author={Gopalswamy, N and Yashiro, S and Michalek, G and Stenborg, G and Vourlidas, A and Freeland, S and Howard, R},
  journal={Earth, Moon, and Planets},
  volume={104},
  pages={295--313},
  year={2009},
  doi={10.1007/s11038-008-9282-7},
  publisher={Springer}
}

@article{harvey1996global,
  title={{The global oscillation network group (GONG) project}},
  author={Harvey, JW and Hill, F and Hubbard, RP and Kennedy, JR and Leibacher, JW and Pintar, JA and Gilman, PA and Noyes, RW and Title, AM and Toomre, J and others},
  journal={Science},
  volume={272},
  number={5266},
  pages={1284--1286},
  year={1996},
  doi={10.1126/science.272.5266.1284},
  url={https://doi.org/10.1126/science.272.5266.1284},
  publisher={American Association for the Advancement of Science}
}

@article{hill2018global,
  title={{The Global Oscillation Network Group Facility—an Example of Research to Operations in Space Weather}},
  author={Hill, Frank},
  journal={Space Weather},
  volume={16},
  number={10},
  pages={1488--1497},
  year={2018},
  doi={10.1029/2018SW002001},
  publisher={Wiley Online Library}
}

@article{jin2017data,
  title={Data-constrained coronal mass ejections in a global magnetohydrodynamics model},
  author={Jin, M and Manchester, WB and van der Holst, B and Sokolov, I and T{\'o}th, G and Mullinix, RE and Taktakishvili, A and Chulaki, A and Gombosi, TI},
  journal={\apj},
  volume={834},
  number={2},
  pages={173},
  year={2017},
  doi={10.3847/1538-4357/834/2/173},
  publisher={IOP Publishing}
}

@article{liu2025physics,
  title={{Physics-based Simulation of the 2013 April 11 Solar Energetic Particle Event}},
  author={Liu, Weihao and Sokolov, Igor V and Zhao, Lulu and Gombosi, Tamas I and Sachdeva, Nishtha and Chen, Xiaohang and T{\'o}th, G{\'a}bor and Lario, David and Manchester, Ward and Whitman, Kathryn and Cohen, Christina M S and Bruno, Alessandro and Mays, M Leila and Bain, Hazel M},
  journal={\apj},
  volume={985},
  number={1},
  pages={82},
  year={2025},
  doi={10.3847/1538-4357/adc4e3},
  publisher={IOP Publishing}
}

@article{liu2026testbed,
  title={Simulated operational testing of the prototype implementation of the SOFIE model: The 2025 space weather prediction testbed exercise},
  author={Liu, Weihao and Zhao, Lulu and Sokolov, Igor V and Whitman, Kathryn and Gombosi, Tamas I and Sachdeva, Nishtha and Adamson, Eric T and Bain, Hazel M and Corti, Claudio and Mays, M Leila and others},
  journal={Space Weather},
  volume={24},
  number={3},
  pages={e2025SW004811},
  year={2026},
  doi={10.1029/2025SW004811},
  publisher={Wiley Online Library}
}

@article{liu2026simulating,
  title={Simulating the Solar Corona with Multiple Solar Photospheric Magnetic Maps during the 2024 April 8 Total Solar Eclipse},
  author={Liu, Xianyu and Liu, Weihao and Manchester IV, Ward B and Welling, Daniel T and T{\'o}th, G{\'a}bor and Gombosi, Tamas I and DeRosa, Marc L and Bertello, Luca and Pevtsov, Alexei A and Pevtsov, Alexander A and others},
  journal={\apj},
  volume={997},
  number={2},
  pages={243},
  year={2026},
  doi={10.3847/1538-4357/ae290f},
  publisher={The American Astronomical Society}
}

@article{maguire2020evolution,
  title={Evolution of the Alfv{\'e}n Mach number associated with a coronal mass ejection shock},
  author={Maguire, Ciara A and Carley, Eoin P and McCauley, Joseph and Gallagher, Peter T},
  journal={\aap},
  volume={633},
  pages={A56},
  year={2020},
  doi={10.1051/0004-6361/201936449},
  publisher={EDP Sciences}
}

@article{manchester2005coronal,
  title={Coronal mass ejection shock and sheath structures relevant to particle acceleration},
  author={Manchester, Ward and Gombosi, TI and De Zeeuw, DL and Sokolov, IV and Roussev, II and Powell, KG and K{\'o}ta, J and T{\'o}th, G and Zurbuchen, TH},
  journal={\apj},
  volume={622},
  number={2},
  pages={1225},
  year={2005},
  doi={10.1086/427768},
  publisher={IOP Publishing}
}

@article{manchester2025high,
  title={High-resolution Simulation of Coronal Mass Ejection--Corotating Interaction Region Interactions: Mesoscale Solar Wind Structure Formation Observable by the SWIFT Constellation},
  author={Manchester, WB and Sachdeva, Nishtha and Kilpua, Emilia and Ala-Lahti, Matti and Soni, Shirsh Lata and Huang, Zhenguang and Chen, Hongfan and Jivani, Aniket and van der Holst, Bart and Szabo, Adam and others},
  journal={\apj},
  volume={992},
  number={1},
  pages={51},
  year={2025},
  doi={10.3847/1538-4357/adf855},
  publisher={IOP Publishing}
}

@article{reiss2023progress,
  title={Progress and challenges in understanding the ambient solar magnetic field, heating, and spectral irradiance},
  author={Reiss, Martin A and Arge, Charles N and Henney, Carl J and Klimchuk, James A and Linker, Jon A and Muglach, Karin and Pevtsov, Alexei A and Pinto, Rui F and Schonfeld, Samuel J},
  journal={Advances in Space Research},
  year={2023},
  doi={10.1016/j.asr.2023.08.039},
  publisher={Elsevier}
}

@article{sachdeva2019validation,
  title={Validation of the Alfv{\'e}n wave solar atmosphere model (AWSoM) with observations from the low corona to 1 Au},
  author={Sachdeva, Nishtha and van Der Holst, Bart and Manchester, Ward and T{\'o}th, Gabor and Chen, Yuxi and Lloveras, Diego G and V{\'a}squez, Alberto M and Lamy, Philippe and Wojak, Julien and Jackson, Bernard V and others},
  journal={\apj},
  volume={887},
  number={1},
  pages={83},
  year={2019},
  doi={10.3847/1538-4357/ab4f5e},
  publisher={IOP Publishing}
}

@article{shi2022awsom,
  title={AWSoM Magnetohydrodynamic Simulation of a Solar Active Region with Realistic Spectral Synthesis},
  author={Shi, Tong and Manchester, Ward and Landi, Enrico and van der Holst, Bart and Szente, Judit and Chen, Yuxi and T{\'o}th, G{\'a}bor and Bertello, Luca and Pevtsov, Alexander},
  journal={\apj},
  volume={928},
  number={1},
  pages={34},
  year={2022},
  doi={10.3847/1538-4357/ac52ab},
  publisher={IOP Publishing}
}

@article{sokolov2013magnetohydrodynamic,
  title={{Magnetohydrodynamic Waves and Coronal Heating: Unifying Empirical and MHD Turbulence Models}},
  author={Sokolov, Igor V and van der Holst, Bart and Oran, Rona and Downs, Cooper and Roussev, Ilia I and Jin, Meng and Manchester, Ward and Evans, Rebekah M and Gombosi, Tamas I},
  journal={\apj},
  volume={764},
  number={1},
  pages={23},
  year={2013},
  doi={10.1088/0004-637X/764/1/23},
  publisher={IOP Publishing}
}

@article{sokolov2021threaded,
  title={{Threaded-field-line Model for the Low Solar Corona Powered by the Alfv{\'e}n Wave Turbulence}},
  author={Sokolov, Igor V and van der Holst, Bart and Manchester, Ward and Ozturk, Doga Can Su and Szente, Judit and Taktakishvili, Aleksandre and T{\'o}th, G{\'a}bor and Jin, Meng and Gombosi, Tamas I},
  journal={\apj},
  volume={908},
  number={2},
  pages={172},
  year={2021},
  doi={10.3847/1538-4357/abc000},
  publisher={IOP Publishing}
}

@inproceedings{stanzione2020frontera,
    author = {Stanzione, Dan and West, John and Evans, R. Todd and Minyard, Tommy and Ghattas, Omar and Panda, Dhabaleswar K.},
    title = {{Frontera: The Evolution of Leadership Computing at the National Science Foundation}},
    year = {2020},
    isbn = {9781450366892},
    publisher = {Association for Computing Machinery},
    doi = {10.1145/3311790.3396656},
    booktitle = {Practice and Experience in Advanced Research Computing 2020: Catch the Wave},
    pages = {106–111},
    numpages = {6}
}

@article{van2014alfven,
  title={{Alfv{\'e}n wave solar model (AWSoM): coronal heating}},
  author={van der Holst, Bart and Sokolov, Igor V and Meng, Xing and Jin, Meng and Manchester, Ward and T{\'o}th, G{\'a}bor and Gombosi, Tamas I},
  journal={\apj},
  volume={782},
  number={2},
  pages={81},
  year={2014},
  doi={10.1088/0004-637X/782/2/81},
  publisher={IOP Publishing}
}

@article{zhao2024solar,
  title={{Solar Wind With Field Lines and Energetic Particles (SOFIE) Model: Application to Historical Solar Energetic Particle Events}},
  author={Zhao, Lulu and Sokolov, Igor and Gombosi, Tamas and Lario, David and Whitman, Kathryn and Huang, Zhenguang and T{\'o}th, G{\'a}bor and Manchester, Ward and van der Holst, Bart and Sachdeva, Nishtha and Liu, Weihao},
  journal={Space Weather},
  volume={22},
  number={9},
  pages={e2023SW003729},
  year={2024},
  doi={10.1029/2023SW003729},
  publisher={Wiley Online Library}
}

@ARTICLE{kong2020quasiperpendicular,
       author = {{Kong}, Fang-Jun and {Qin}, Gang},
        title = "{Suprathermal Electron Acceleration by a Quasi-perpendicular Shock: Simulations and Observations}",
      journal = {\apj},
         year = 2020,
        month = jun,
       volume = {896},
       number = {1},
          eid = {136},
        pages = {136},
          doi = {10.3847/1538-4357/ab8e32},
       adsurl = {https://ui.adsabs.harvard.edu/abs/2020ApJ...896..136K}
}

@article{darnel2022suvi,
  title={The GOES-R solar UltraViolet imager},
  author={Darnel, Jonathan M and Seaton, Daniel B and Bethge, Christian and Rachmeler, Laurel and Jarvis, Alison and Hill, Steven M and Peck, Courtney L and Hughes, J Marcus and Shapiro, Jason and Riley, Allyssa and others},
  journal={Space Weather},
  volume={20},
  number={4},
  pages={e2022SW003044},
  year={2022},
  doi={10.1029/2022SW003044},
  publisher={Wiley Online Library}
}

@ARTICLE{zucca2014understanding,
       author = {{Zucca}, P. and {Pick}, M. and {D{\'e}moulin}, P. and {Kerdraon}, A. and {Lecacheux}, A. and {Gallagher}, P.~T.},
        title = "{Understanding Coronal Mass Ejections and Associated Shocks in the Solar Corona by Merging Multiwavelength Observations}",
      journal = {\apj},
         year = 2014,
        month = nov,
       volume = {795},
       number = {1},
          eid = {68},
        pages = {68},
          doi = {10.1088/0004-637X/795/1/68},
archivePrefix = {arXiv},
       eprint = {1409.3691},
 primaryClass = {astro-ph.SR},
       adsurl = {https://ui.adsabs.harvard.edu/abs/2014ApJ...795...68Z}
}

@article{liu2026counterintuitive,
  title={Counterintuitive Magnetic Connectivity and Energetic Particle Flux Differences among Nearby Spacecraft During the 2023 February 24 Solar Energetic Particle Event},
  author={Liu, Weihao and Liu, Xianyu and Lario, David and Zhao, Lulu and Gombosi, Tamas I and Shane, Alexander D and Sokolov, Igor V},
  journal={arXiv preprint arXiv:2606.02445},
  doi={10.48550/arXiv.2606.02445},
  year={2026}
}

@article{hegedus2021tracking,
  title={Tracking the Source of Solar Type II Bursts through Comparisons of Simulations and Radio Data},
  author={Hegedus, Alexander M and Manchester IV, Ward B and Kasper, Justin C},
  journal={\apj},
  volume={922},
  number={2},
  pages={203},
  year={2021},
  doi={10.3847/1538-4357/ac2361},
  publisher={The American Astronomical Society}
}
\bibliographystyle{aasjournalv7}



\end{document}